\documentclass[11pt,a4paper]{article}
\pdfoutput=1

\usepackage{amsmath}
\usepackage{amssymb}
\usepackage{cite}
\usepackage{graphicx}
\usepackage{bbm}
\usepackage{color}
\usepackage{subcaption}

\newcommand{\be}{\begin{equation}}  
\newcommand{\ee}{\end{equation}}  
\newcommand{\ol}[1]{\overline{#1}}
\newcommand{\hc}{+\,\mathrm{h.c.}}

\newcommand{\SU}[1]{\ensuremath{\mathrm{SU}(#1)}}
\newcommand{\U}[1]{\ensuremath{\mathrm{U}(#1)}}
\newcommand{\into}{\ensuremath{\,\rightarrow\,}}

\newcommand{\tensor}{\ensuremath{\otimes}}

\newcommand{\wt}[1]{\widetilde{#1}}

\renewcommand{\sb}{\ensuremath{s_\beta}}
\newcommand{\stb}{\ensuremath{s_{2\beta}}}
\newcommand{\cb}{\ensuremath{c_\beta}}

\newcommand{\wh}[1]{\widehat{#1}}

\title{
\vspace{-4.5ex}
{\normalsize \raggedright
February 2017\\[10ex]
}
\textbf{Well-tempered n-plet dark matter} 
\vspace{2ex}
}

\author{\large A.~Bharucha$^a$, F.~Br\"ummer$^{b}$ and R.~Ruffault$^{b}$\\[1ex]
\textit{\normalsize $^a$ CPT, UMR7332, CNRS and Aix-Marseille Universit\'e and Universit\'e de Toulon}\\
\textit{\normalsize 13288 Marseille, France}\\
\textit{\normalsize $^b$ LUPM, UMR5299, Universit\'e de Montpellier and CNRS}\\
\textit{\normalsize 34095 Montpellier, France}\\[1ex]
\vspace{3ex}
}

\date{}

\begin{document}

\maketitle

\begin{abstract} \noindent
We study simple effective models of fermionic WIMP dark matter, where the dark matter candidate is a mixture of a Standard Model singlet and an $n$-plet of $\SU{2}_L$ with $n\geq 3$, stabilized by a discrete symmetry. The dark matter mass is assumed to be around the electroweak scale, and the mixing is generated by higher-dimensional operators, with a cutoff scale $\gtrsim$ TeV. For appropriate values of the mass parameters and the mixing we find that the observed dark matter relic density can be generated by coannihilation. Direct detection experiments have already excluded large parts of the parameter space, and the next-generation experiments will further constrain these models. 
\end{abstract}

\section{Introduction}

The WIMP paradigm is based on the observation that the thermal relic density of a stable, electrically neutral particle with electroweak cross-section and electroweak-scale mass roughly corresponds to the observed dark matter abundance, $\Omega h^2=0.1199\pm0.0022$~\cite{planck}. However, a closer look reveals that this correspondence is quantitatively not very precise. In fact, adding a new particle with suitable $\SU{2}\times\U{1}$ quantum numbers to the Standard Model, one finds that its mass must be of the order of $1-10$ TeV, i.e.~$1-2$ orders of magnitude larger than the electroweak scale, if its thermal relic density is to reproduce the observed value \cite{Cirelli:2005uq}. A thermal relic dark matter candidate with a mass ${\cal O}(100$ GeV$)$ must necessarily originate mostly from an $\SU{2}\times\U{1}$ singlet, and feel the electroweak gauge interactions at most through a small coupling to a non-singlet (the prototypical example being the ``well-tempered neutralino'' \cite{ArkaniHamed:2006mb} of the MSSM, which is mostly a bino).

While minimal extensions of the Standard Model with a WIMP dark matter particle at ${\cal O}(10$ TeV$)$ are of interest on their own, they have at least two disadvantages. First, conceptually, it is more difficult to envisage how such heavy states could naturally emerge from some extension of the Standard Model which solves the electroweak hierarchy problem. And second, practically, they are difficult to directly test at colliders: For electroweak production, the reach of the LHC is limited to masses of at most a few hundred GeV before the cross-section becomes negligibly small. Similar arguments apply for models of mixed WIMP dark matter without singlets.\footnote{See e.g.~\cite{Dedes:2014hga} and \cite{Tait:2016qbg} for recent work on non-supersymmetric mixed WIMP models.} 

In this article we therefore investigate the case where the dark matter candidate is mostly an $\SU{2}\times\U{1}$ singlet but has a small mixing with another state charged under $\SU{2}\times\U{1}$. Our models can be regarded as generalizations of the well-tempered neutralino scenario with a bino-like LSP (see e.g.~\cite{ArkaniHamed:2006mb}, and \cite{Bramante:2015una, Baer:2016ucr,Beneke:2016jpw,Han:2016qtc,Ellwanger:2016sur,Huang:2017kdh,Badziak:2017the,Chakraborti:2017dpu} and references therein for some recent studies) to non-supersymmetric settings and allowing for more exotic electroweak representations. 

In detail, we consider a minimal extension of the Standard Model by a fermionic gauge singlet $\chi$ and a fermion $\psi$ transforming in the $n$-dimensional representation ${\bf n}_Y$ of $\SU{2}\times\U 1$. Odd-dimensional representations are real, and the model is free of anomalies for a hypercharge $Y=0$. Even-dimensional representations require us to add a Dirac partner $\ol\psi$ for $\psi$ transforming in the $\ol{\bf n}_{-Y}$. We further impose a $\mathbb{Z}_2$ symmetry under which the new particles are odd while the Standard Model particles are even; this forbids any mixing with the Standard Model leptons and ensures the stability of the lightest mass eigenstate. We give a Majorana mass (for $n$ odd) or a Dirac mass (for $n$ even) of the order of the electroweak scale to $\psi$, and a somewhat smaller Majorana mass to $\chi$.

The $n=2$ and $n=3$ cases are familiar from supersymmetry (corresponding, respectively, to a higgsino-bino-like and to a wino-bino-like neutralino as the dark matter candidate with all other superpartners heavy). Qualitatively new effects appear starting with $n=4$. Notably, in that case the spectrum contains multiply charged states, which opens up new possibilities for testing these models at colliders: the production cross-section can be sizeable, and their decay length is large which may lead to exotic signatures in the detector. We will investigate the collider physics of our models in detail in a future publication \cite{BBDR}, and for now concentrate on their dark matter properties. 

Specifically, we study the representations ${\bf n}_Y={\bf 3}_0,\,{\bf 4}_{\frac{1}{2}}$, and ${\bf 5}_0$ in some detail. In all these models the dark matter candidate (composed mainly of $\chi$) mixes with the $n$-plet $\psi$ via a higher-dimensional operator. This mixing generates the appropriate thermal relic density. We remain agnostic about the UV completion and about the origin of the mixing operator, and only study the resulting phenomenology. A dimension-5 coupling of $\chi$ to the Higgs bilinear could in principle also influence the relic density, but we find that direct detection bounds constrain the associated Wilson coefficient so severely that its contribution to the annihilation cross section is negligible. Our models will be further tested with the next generation of direct detection experiments.

In the following section, we will present these models in more detail. We will then proceed in Sec.~\ref{sec:DM} to discuss the dark matter properties, i.e.~the relic density and direct detection cross section. We will finally present our numerical results and constraints on these models, as well as the future prospects in Sec.~\ref{sec:results}, and conclude in Sec.~\ref{sec:conclusions}. Some technical details are relegated to the appendix.

\section{Models}
\label{sec:Models}
For $n$ odd, and specifically $n=3$ and $n=5$, the Lagrangian of our model is
\be\label{eq:LDM-odd}
{\cal L_{\rm DM}}=i\,\psi^\dag\ol\sigma^\mu D_\mu\psi+i\,\chi^\dag\ol\sigma^\mu\partial_\mu\chi-\left(\frac{1}{2} M\psi\psi+\frac{1}{2}m\chi\chi\hc\right)+{\cal L}_{\rm quartic}+{\cal L}_{\rm mix}\,,
\ee
where
\be\label{eq:Lquartic}
{\cal L}_{\rm quartic}=\frac{1}{2}\frac{\kappa}{\Lambda}\phi^\dag\phi \chi\chi \hc
\ee
and, schematically,
\be\label{eq:Lmix-odd}
{\cal L}_{\rm mix}=\frac{\lambda}{\Lambda^{n-2}}(\phi^\dagger \phi)^\frac{n-1}{2} \psi\chi\hc
\ee
Here $\psi$ is a Majorana fermion transforming in the ${\bf n}_0$  of $\SU{2}_L\times\U 1_Y$, $\chi$ is a Majorana singlet, and $\phi$ is the Standard Model Higgs doublet. 

For $n$ even, in particular $n=2$ or $n=4$, the Lagrangian is
\be\label{eq:LDM-even}
{\cal L_{\rm DM}}=i\,\psi^\dag\ol\sigma^\mu D_\mu \psi+i\,\ol\psi^{\dag}\ol\sigma^\mu D_\mu\ol\psi+i\,\chi^\dag\ol\sigma^\mu\partial_\mu\chi-\left( M\psi\ol\psi+\frac{1}{2}m\chi\chi\hc\right)+{\cal L}_{\rm quartic}+{\cal L}_{\rm mix}\,,
\ee
where ${\cal L}_{\rm quartic}$ is given by Eq.~\eqref{eq:Lquartic} as before, ${\cal L}_{\rm mix}$ is (again schematically),
\be\label{eq:Lmix-even}
{\cal L}_{\rm mix}=\frac{1}{\Lambda^{n-2}}(\phi^\dagger \phi)^\frac{n-2}{2}\left(\lambda\; \phi \chi\psi -\lambda'\;\phi^\dag\chi\ol\psi\hc\right)
\ee
and $(\psi,\ol\psi^{\dag})$ form a Dirac spinor transforming in the ${\bf n}_{\frac{1}{2}}$. All new fermions are odd under a global $\mathbbm{Z}_2$.

In order to obtain the observed relic density with electroweak-scale masses, the lightest neutral mass eigenstate should be $\chi$-like. However we allow for a small mass mixing ${\cal L}_{\rm mix}$ between $\chi$ and the electrically neutral components of $\psi$ after electroweak symmetry breaking. For $n>2$ this is due to a higher-dimensional operator, so ${\cal L}_{\rm DM}$ is an effective Lagrangian valid up to the scale $\Lambda$, around which additional states appear in the spectrum. We will assume that $\Lambda$ is sufficiently large for these new states to play essentially no role at electroweak energies, except to induce the higher-dimensional operators in Eqs.~\eqref{eq:LDM-odd} or ~\eqref{eq:LDM-even}. This is already the case for $\Lambda\sim$ TeV if the new physics is weakly coupled, $\lambda,\lambda',\kappa \lesssim 1$. The operator ${\cal O}_{\rm quartic}$ is the leading operator allowing for direct $\chi\chi$ annihilation into Standard Model states, without involving $\psi$. It can significantly influence the dark matter properties of the model, given that it is of dimension 5 while the mixing operators ${\cal L}_{\rm mix}$ are of dimension greater than $5$ for $n>3$.
 
The dimension-5 operators
\begin{align}
&\frac{1}{2}\frac{\kappa'}{\Lambda}\phi^\dag\phi \psi\psi\hc\qquad&(n\text{ odd})\label{eq:dim5-1}\\
 &\frac{\kappa'}{\Lambda}\phi^{\dag}\phi \psi\ol\psi\hc\qquad&(n\text{ even})\label{eq:dim5-2}\\
&\frac{\zeta_1}{\Lambda}(\phi^\dag\tau^a\phi)(\ol\psi t^a\psi)
-\frac{\zeta_2}{\Lambda}(\phi_i \tau^a{}^i_j\phi^j)(\ol\psi_I t^a{}_J^I\ol\psi^J)
-\frac{\zeta_3}{\Lambda}(\phi^\dag_i\tau^a{}^i_j\phi^{\dag j})(\psi_I t^a{}_J^I\psi^J)\hc\qquad&(n\text{ even})\label{eq:dim5-3}
\end{align}
(with $\tau^a$ generating the ${\bf 2}$ and $t^a$ generating the ${\bf n}$) will have an impact on the mass spectrum after electroweak symmetry breaking, and thus indirectly affect the $\chi$ relic density. While $\kappa'$ can always be set to zero by a redefinition of $M$, the mass shift induced by $\zeta_{1,2,3}$ differs between charged and neutral mass eigenstates and will therefore need to be taken into account.

This list of higher-dimensional operators is far from exhaustive, even at dimension 5 or 6.\footnote{For a classification of dimension-6 operators coupling a parity-stabilized singlet dark matter sector to the Standard Model, see \cite{Duch:2014xda}. For the case of electroweak doublets, see \cite{Dedes:2016odh}.} However, we will restrict our analysis to the operators we have listed above, for the following reasons. First of all, we assume that dimension-6 couplings between the dark matter candidate and the SM fermions, such as $(q_L\chi) (q_L^\dag\chi^\dag)/\Lambda^2$, are suppressed, since these would otherwise lead to unacceptably large flavour-changing neutral currents. Moreover, for the observables we are interested in (the mass spectrum, the thermal relic density and the direct and indirect detection cross sections) any higher-derivative couplings play at most a subdominant role. Finally, we can neglect any subleading couplings which affect our observables only through singlet-$n$-plet mixing.

\subsection{``${\bf n=0}$'': A pure singlet?}\label{puresinglet}
Since the dimension-5 operator ${\cal L}_{\rm quartic}$ of Eq.~\eqref{eq:Lquartic} allows for direct $\chi$ annihilation into SM states, can we simply build a more minimal dark matter model without any $n$-plet $\psi$? In other words, can we reproduce the observed dark matter relic density simply with a single electroweak-scale singlet $\chi$, with all other states substantially heavier (e.g.~with masses $\Lambda\gtrsim$ TeV) such that they should be integrated out at low energies, thereby inducing the coupling $\kappa$? 

For a sufficiently large $\kappa$ (or equivalently a sufficiently low suppression scale $\Lambda$), ${\cal L}_{\rm quartic}$  can indeed lead to the correct relic density via thermal freeze-out. However, such large values of $\kappa$ are by now excluded by direct detection. We will present some more details in Sec.~\ref{sec:results}, but given that this scenario is of no phenomenological interest, our main focus will be on models which contain an $n$-plet along with the singlet.

\subsection{$\bf{n=2}$: The well-tempered higgsino-bino and its non-SUSY generalisation}

The case $n=2$ is familiar from the MSSM: the $\mathbbm{Z}_2$ symmetry corresponds to $R$-parity, $\chi$ to the bino and $(\psi,\ol\psi)$ to the higgsinos. The wino is effectively decoupled, $M_2\gg M_1,\,\mu$. Likewise, the squark, slepton and non-standard Higgs boson masses are large compared to $\mu$ and $M_1$. Dark matter is the lightest neutralino. In the $n=2$ case the model is renormalizable, because gauge invariance allows for a bino-higgsino-Higgs Yukawa coupling. Since this system has been extensively studied both in the supersymmetric (where it is excluded for scenarios giving the correct relic density, see e.g.~\cite{Barducci:2015ffa}) and in the non-supersymmetric context, we merely refer to the literature \cite{Mahbubani:2005pt, ArkaniHamed:2006mb, Cohen:2011ec,Yanagida:2013uka,Cheung:2013dua,Abe:2014gua,Bramante:2014tba,Calibbi:2015nha,Freitas:2015hsa,Badziak:2015qca,Barducci:2015ffa,Yaguna:2015mva,Horiuchi:2016tqw,Banerjee:2016hsk,Basirnia:2016szw,Kearney:2016rng,Huang:2017kdh,Abe:2017glm}.

\subsection{$\bf{n=3}$: The well-tempered wino-bino and its non-SUSY generalisation}

The case $n=3$ can also appear in the MSSM when we identify $\Lambda=\mu$, $\lambda= gg'\sin(2\beta)$ and $\kappa={g'}^2\sin(2\beta)$ (or more precisely $\lambda=\tilde g_u\tilde g_d'+\tilde g_d \tilde g_u'$ and $\kappa=2\tilde g_u'\tilde g_d'$ \cite{Giudice:2004tc} in the ``split SUSY'' case of a parametrically large SUSY breaking scale). Here the lightest neutralino which constitutes dark matter is a mixture of mostly wino and bino (with necessarily some higgsino component as well). Wino-bino mixing is forbidden by gauge invariance at the renormalizable level, but the mixing term $\mathcal{L}_{\rm mix}$ introduced in Eq.~\eqref{eq:Lmix-odd} is generated by integrating out the higgsinos. Some of the technical details are recapitulated in Appendix A.

Although this example has also been extensively studied before (see e.g.~\cite{ArkaniHamed:2006mb,Arina:2014xya,Harigaya:2014dwa, Nagata:2015pra, Rolbiecki:2015gsa}), we will investigate it in some detail in order to pave the ground for our later analysis of even higher representations. Moreover, the fact that for this case a simple and well-studied explicit UV completion is available in the MSSM, allows for some useful checks and comparisons of the effective theory with the complete one. The mixing term is
\be
\begin{split}
 {\cal L}_{\rm mix}=\frac{\lambda}{\Lambda}\phi^\dag\tau^a\phi\;\psi^a\chi\hc
\end{split}
\ee
where $\tau^a=\sigma^a/2$. There is one charged mass eigenstate originating from $\psi$, and two neutral ones which are superpositions of $\psi^3$ and $\chi$. After absorbing the mass shifts proportional to $\kappa$ and $\kappa'$ (see Eqs.~\eqref{eq:Lquartic} and \eqref{eq:dim5-1}) into $M$ and $m$, the $\psi^3-\chi$ mixing angle is
\be\label{eq:theta3}
\theta\approx\frac{\sqrt{2}\lambda v^2}{\Lambda(M-m)}
\ee
to leading order in $v/\Lambda$, where $v=174$ GeV is the electroweak vev. This expansion breaks down at the mass-degenerate point $M=m$; we are therefore implicitly assuming that the mass difference between the $n$-plet and the singlet is not parametrically smaller than the electroweak scale. (We will see that $(M-m)\sim $ few $\cdot\, 10$ GeV for the cases of interest.) Moreover, Eq.~\eqref{eq:theta3} may be a poor approximation to the true mixing angle if the coupling $\lambda$ is accidentally so small that the higher-order terms in the $v/\Lambda$ expansion dominate. This is e.g.~the case in the bino-wino scenario of the MSSM if either $\mu$ is small or $\tan\beta$ is large, roughly for $\mu\lesssim m_Z\tan\beta$.

\subsection{$\bf{n=4}$: The well-tempered quadruplet-singlet}

Even-dimensional representations are slightly more complicated because they are no longer strictly real. We will discuss the example of ${\bf n}_Y={\bf 4}_{\frac{1}{2}}$ or the quadruplet-singlet model. The mixing term is
\be
{\cal L}_{\rm mix}=\left(\frac{\lambda}{\Lambda^2}\;\epsilon_{jl}\epsilon_{km}d_I^{ijk}\;\phi^{\dag}_i\phi^l\phi^m\,\chi\ol\psi^I-\frac{\lambda'}{\Lambda^2}\;\epsilon_{kl}d_I^{ijk}\;\phi^{\dag}_i\phi^{\dag}_j\phi^{l}\chi\psi^I\hc\right)\,\,.
\ee
The notation and some more technical details are explained in App.~B. The spectrum now consists of a doubly charged Dirac particle $\chi^{\pm\pm}$, two singly charged Dirac particles $\chi^{\pm}_{1,2}$, and three neutral Majorana particles $\chi^0_{1,2,3}$. The dark matter candidate $\chi^0_1$ is still mostly $\chi$-like by assumption, but now contains small admixtures from both of the two neutral states contained in $\psi$ and $\ol\psi$, hence there are two potentially relevant mixing angles $\theta_+$ and $\theta_-$. At leading order in $v/\Lambda$ these are given by
\be\label{eq:theta4}
\theta_\pm\approx\frac{1}{\sqrt{6}}\frac{(\lambda\pm\lambda')v^3}{\Lambda^2(M\mp m)}\,.
\ee
For the validity of this approximation, the same comments apply as in the triplet case.
We have once more absorbed the mass shifts due to electroweak symmetry breaking, see Eqs.~\eqref{eq:Lquartic}, \eqref{eq:dim5-2} and \eqref{eq:dim5-3}, into $M$ and $m$. Note that the operators of Eq.~\eqref{eq:dim5-3} induces a mass splitting which is not $\SU{2}$ invariant, hence in the presence of a nonzero $\zeta_{1,2,3}$ the tree-level masses of the charged states will be different from $M$.

\subsection{${\bf n}={\bf 5}$: The well-tempered quintuplet-singlet}

After the triplet-singlet model, the simplest case for odd $n$ is the quintuplet-singlet model, ${\bf n}_Y={\bf 5}_0$. The mixing term is
\be
\begin{split}
 {\cal L}_{\rm mix}=\frac{\lambda}{\Lambda^3}\;C_{A\,ik}^{j\ell}\;\phi^{\dag i}\phi_j\phi^{\dag k}\phi_\ell \psi^A \chi\hc
\end{split}
\ee
The explicit form of the $\SU{2}$ tensor  $C_{A\,ik}^{j\ell}$, along with more technical details, is given in App.~C. There are again two neutral mass eigenstates, superpositions of $\psi^5$ and $\chi$, as well as a singly-charged and a doubly-charged mass eigenstate emerging from $\psi$. After absorbing the mass shifts proportional to $\kappa$ and $\kappa'$ into $M$ and $m$, the mixing angle is, to leading order in $v/\Lambda$,
\be\label{eq:theta5}
\theta\approx\sqrt{\frac{2}{3}}\frac{\lambda v^4}{\Lambda^3(M-m)}\,.
\ee

\section{Dark matter properties}
\label{sec:DM}
In accordance with the usual convention for supersymmetric neutralinos, we denote by $\chi^0_i$ the neutral mass eigenstates, ordered by their masses. As the singlet component of the lightest neutral $\mathbb{Z}_2$-odd particle $\chi^0_1$ in our model does not couple to the SM, the dark matter properties  depends strongly both on the coupling of the $n$-plet to the Higgs boson, $\kappa$ and on the mixing angle between the singlet and $n$-plet. It further depends on the mass parameters of the singlet and $n$-plet. We are interested in masses of the order of the electroweak scale, with the singlet mass being smallest, such that $\chi^0_1$ is dominantly singlet-like.

\subsection{The relic density}
\label{sec:RD}
Since the mixing between the $\chi^0_i$ is due to higher-dimensional operators, it can in principle be arbitrarily small if the associated new physics scale is large. Supposing that it is the mixing term ${\cal L}_{\rm mix}$ rather than the direct annihilation term ${\cal L}_{\rm quartic}$ which dominates the annihilation cross section, the following distinct cases may arise:
\begin{enumerate}
 \item If the $n$-plet-singlet mixing angle $\theta$ is too small for $\chi^0_1$ to ever be in equilibrium at some point during the thermal history of the universe, then the dark matter relic density can only be predicted if some initial conditions are assumed. For example, one can envisage a scenario where the initial $\chi^0_1$ abundance is zero, and subsequently grows due to ``freeze-in''-like processes involving tiny couplings to the SM. For certain cases, this is independent of the UV physics, e.g.~the reheating mechanism, and therefore calculable without making further assumptions. We comment more on the freeze-in mechanism in Sec.~\ref{sec:freeze-in}. For this case there can also be a freeze-out contribution to the relic density from the freeze-out of $\chi^0_2$ and its subsequent decay to $\chi^0_1$.

 \item A second case is that of mixing angles $\theta$ which are small but nevertheless large enough to allow for rapid decays $\psi_i\into\chi^0_1\,+$ SM and scattering processes $\chi^0_1+{\rm SM}\into\psi_i\,+$ SM (where $\psi_i$ stands for any of the $n$-plet-like particles, for example $\chi^0_2$) through the mixing operator ${\cal L}_{\rm mix}$. The  $\chi^0_1$ abundance is then completely determined by $\psi_i$ annihilation, which in turn depletes the $\chi^0_1$ number density because $\chi^0_1$ remains in equilibrium with $\psi_i$. As noted e.g.~in \cite{ArkaniHamed:2006mb}, the dark matter relic density therefore becomes effectively independent of the precise value of $\theta$ and depends only on the masses. Even for very small mixing angles one may reproduce the observed value $\Omega h^2=0.1199\pm0.0022$~\cite{planck} by choosing the mass splitting accordingly. Note that this case is somewhat fine-tuned: Even though a relative mass difference of the order of $10\%$ does not seem too unnatural, the relic density is very sensitive to its exact value, and therefore a very precise parameter choice is needed to match observation.
 
 \item For somewhat larger mixing angles $\theta\gtrsim 0.1$, annihilation processes directly involving $\chi^0_1$ can become important. In that case the relic density is no longer exclusively determined by the $n$-plet component of the $\chi^0_1$, and becomes a function of both the masses and $\theta$. For electroweak-scale masses and a mass splitting of the order of 20 GeV it is always possible to find the observed value of the relic density by choosing $\theta$ suitably, because the system transitions between much larger values (without coannihilation for a singlet-like dark matter candidate) and much smaller ones (for an $n$-plet-like dark matter candidate with a large annihilation cross section). This will be confirmed by our numerical analysis, see Figs.~\ref{TSM_Omega}, \ref{QQSM_Omega} and \ref{QSM_Omega}.
 
 \item As the mixing angle becomes ${\cal O}(1)$, our effective theory description is approaching the limit of its validity. In this region the relic density depends on the details of what happens at the new physics scale $\Lambda$, and we can no longer make any reliable model-independent statement. However, since we are interested in electroweak-scale masses, this case is less interesting phenomenologically because an ${\cal O}(1)$ $n$-plet admixture would lead to a much too large annihilation cross section, and hence a much too small relic density to fit observation.
\end{enumerate}

In the presence of the operator ${\cal L}_{\rm quartic}$, the singlet can directly annihilate into SM states without relying on the mixing. However, it turns out that in the freeze-out case, values of $\kappa/\Lambda$ large enough to significantly influence the annihilation cross section are already ruled out by direct detection experiments, discussed later.

\subsection{Freeze-in}
\label{sec:freeze-in}
The freeze-in mechanism for dark matter generation is an alternative to the more common freeze-out, where it is also possible to reproduce the measured relic abundance of the universe without depending on the UV quantities describing the early history of the universe  such as the reheating temperature~\cite{McDonald:2001vt,Hall:2009bx}. 
In this scenario the dark matter candidate interacts very feebly with the thermal bath (generally through renormalizable interactions), and is therefore thermally decoupled. 
One further assumes
the dark matter abundance in the early universe to be negligibly small,  increasing only when it is produced via interactions with the bath, particularly when the temperature drops below the mass of the dark matter particle.
One can calculate the resulting relic abundance for a coupling of the type $\alpha' \phi \psi\chi$ where, by solving the Boltzmann equation for $n_\chi$, the number density of $\chi$ particles takes the approximate form
\begin{equation}
\dot{n}_\chi+3 n_\chi H\approx \frac{g_{\psi}M^2\Gamma_{\psi}}{2\pi^2}T K_1(M/T),
\end{equation}
where $K_1$ is the first modified Bessel function of the second kind, and $\Gamma_{\psi}\sim \alpha^2 M/(8\pi)$ is the decay rate of the $\psi$ into the dark matter particle $\chi$ and the Higgs.
This is solved using $\dot{T}\approx -HT$.
The resulting relic abundance takes the form
\begin{equation}
\Omega_\chi h^2|_{\rm tot}\simeq \frac{1.09\times 10^{27}}{g_*^S\sqrt{g_*^\rho}}\frac{m_X \Gamma_{\psi}}{M^2}
\end{equation}
In the case of non-renormalizable interactions, however, on solving the Boltzmann equation one finds that the relic density depends on the UV physics, in particular the reheating temperature $T_R$. For example, for the operator $\tfrac{\alpha}{\Lambda}\phi^\dagger\phi \psi\chi$ the relic abundance takes the form 
\begin{equation}
\Omega_\chi h^2|_{\rm tot}\propto\frac{\alpha^2 m T_R}{\Lambda^2}.
\end{equation}
Since we have no knowledge of the reheating temperature, models for which the UV freeze-in mechanism prevails are less predictive in terms of the masses and couplings required to obtain the correct relic density, however such scenarios may be more generic than IR freeze-in mechanisms, and several have been studied in ~\cite{Elahi:2014fsa}.

As seen in the Lagrangian given in Eq.~\eqref{eq:LDM-odd}, on expanding about the Higgs vev we find that the couplings of the DM particle are described by two three-field and two four-field operators in the case $n=3$. For the former, the coupling is renormalizable, and the DM production would be independent of the UV physics of the early universe. The latter two however have dimensionful couplings resulting in a dependence on the reheating temperature as shown above. 
Out of the three-field operators, one will contribute to the freeze-in via decays of the triplet to the singlet and the Higgs (provided there is sufficient mass difference), and the other will contribute via $2\to 2$ scattering
One can calculate the value of the reheating temperature below which the renormalizable operators dominate, however this is found to be relatively low, of the order of 10 TeV.
If this condition were to be satisfied, in general, the values of the couplings giving the correct relic density for these models are of $\mathcal{O}(10^{-11} -10^{-13} )$.
We will not consider freeze-in production of dark matter further here, though it offers interesting model-building opportunities.

\subsection{Direct detection }
\label{sec:DD}
 Having discussed the relic density for these models, we now address direct detection constraints. Direct detection experiments probe the interaction between the dark matter particle and nucleons. The spin-independent bounds are more constraining for the parameter space of interest to us than those which are spin-dependent. The spin-independent cross-section $\sigma_{\rm SI}$ is sensitive to the interactions between the $\chi^0_1$ and the nucleons via Higgs exchange. Note that the relative size of the Higgs Yukawa couplings means that the contribution due to the Higgs interacting with strange quarks dominates the result, affecting the total uncertainty of the calculation as the  the strange quark content of the nucleon contributes an uncertainty on the cross section of about 20\%~\cite{Durr:2015dna}, which in combination with the uncertainty on the local relic density of dark matter, $\rho=0.3\pm0.1\, \mathrm{GeV}/\mathrm{cm}^3$~\cite{Bovy:2012tw}, results in weakening the bounds by a factor of two. 
The bounds on $\sigma_{\rm SI}$ were recently updated by both the LUX \cite{Akerib:2016vxi} and the PandaX \cite{Tan:2016zwf} collaborations, of which the former is the most constraining in the mass range we consider. Future results are expected from Xenon1T~\cite{Aprile:2015uzo}.
Note that the exclusion limits provided by the experiments assume that the dark matter particle provides the total relic density as measured by Planck. Therefore in the case of an underabundance of the dark matter particle, the limit $\sigma_{\rm SI}^{\rm exp}$ should be rescaled by
\begin{equation}
\frac{\Omega h^2|_{\rm exp}}
{\Omega h^2|_{\rm \chi}} \sigma_{\rm SI}^{\rm exp}\,
\label{eq:rescaling}
\end{equation}
where $\Omega h^2|_{\rm exp}$ is the value measured by Planck and $\Omega h^2|_{\rm \chi}$ is the thermal relic density of the dark matter candidate in our model. Note that in the plots we instead rescale the theoretical cross section in order to compare with a fixed experimental limit.
\texttt{micrOMEGAs} is used to compute the spin-independent scattering cross section.

Since the coupling to Higgs bosons requires mixing between the single and the $n$-plet,
the limits are sensitive to the parameters affecting the mixing,
\be
\theta \propto \frac{\lambda v^{n-1}}{\Lambda^{n-2}(M-m)}\,,
\ee
as well as the coupling $\kappa$. To be precise, on decreasing $\kappa$, on increasing the splitting or on increasing the scale $\Lambda$ we find that $\sigma_{\rm SI}$ decreases. 
The behaviour of the relic density is approximately the inverse, i.e.~it increases as $\sigma_{\rm SI}$ decreases. Therefore the upper limit on the relic density from Planck acts in a complementary manner to the upper bound on $\sigma_{\rm SI}$, meaning that a region in parameter space may remain exhibiting the correct DM properties. There is however a difference between the relic density and the $\sigma_{\rm SI}$ behaviour: while the latter is independent of the mass of the dark matter particle, the former is approximately inversely proportional to this quantity. 

\section{Results}
\label{sec:results}

We have analysed our models numerically, obtaining the relic density and the spin-independent WIMP-nucleon cross section, using the public code {\tt micrOMEGAs 4.3.1} \cite{Belanger:2014vza, Belanger:2013oya}, with the model files generated with the help of {\tt FeynRules 2.3} \cite{Alloul:2013bka}. We then compare our results with the latest bounds: $\Omega h^2$ from Planck~\cite{planck} and $\sigma_{\rm SI}$ recently updated by LUX \cite{Akerib:2016vxi}.

\begin{figure}[ht]
    \centering
    \includegraphics[width=.5\textwidth]{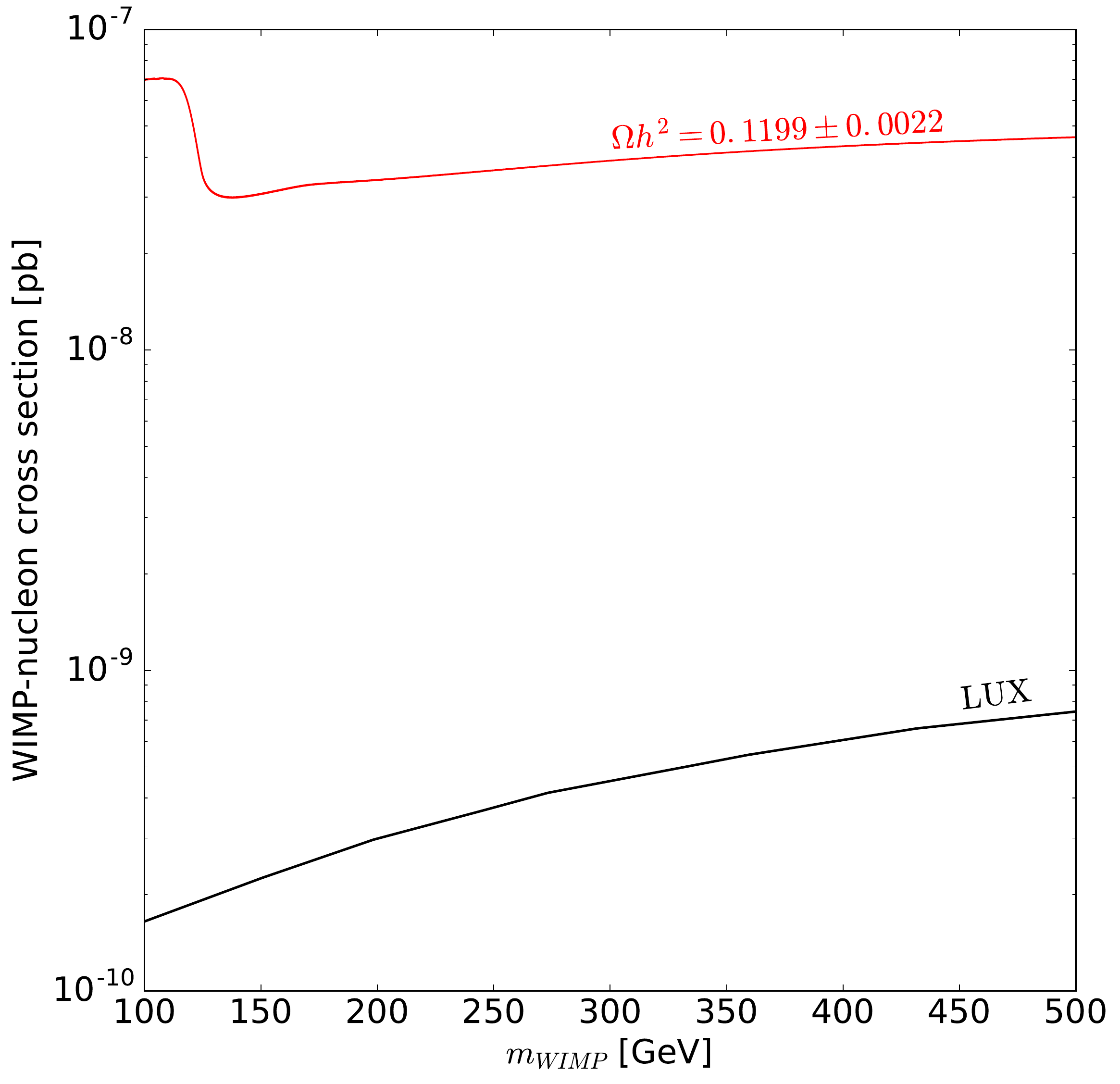}
\caption{For the pure singlet case, with a Wilson coefficient $\kappa$ normalised to $1$ and the scale $\Lambda$ chosen such as to generate the observed relic density via freeze-out,  the corresponding WIMP-nucleon cross section is ruled out by direct detection (we compare  the normalised cross-section with the latest LUX limits \cite{Akerib:2016vxi}).}
\label{SSMfig}
\end{figure}

As already anticipated in Sec.~\ref{puresinglet}, 
a pure singlet $\chi$ annihilating through the dimension-5 operator  of ${\cal L}_{\rm quartic}$ could lead  to the correct relic density via thermal freeze-out if $\kappa/\Lambda$ is sufficiently large. However, such large values of the coupling $\kappa/\Lambda$ are already definitively excluded by direct detection, as Fig.\ref{SSMfig} shows. Thus, our conclusion is that we confirm an electroweak-scale singlet alone not to be compatible with observation; we need the additional electroweak-scale degrees of freedom $\psi$.

We nevertheless point out two potential loopholes to this argument. The first is that we only take the dimension-5 operator ${\cal L}_{\rm quartic}$ into account, neglecting other possible annihilation channels, in particular those into SM fermions via operators such as $(q_L\chi) (q_L^\dag\chi^\dag)/\Lambda^2$. However, operators of this kind are already dimension-6. Moreover they generically need to be suppressed by a much larger scale than a TeV to avoid large flavour-changing neutral currents; neglecting them as we do here therefore corresponds to a well-motivated assumption on the UV completion.\footnote{Unless the sector which generates them carries some special flavour structure, e.g. minimal flavour violation, which is one possibility to indeed leave this loophole open.} The second loophole is that we have assumed the relic density to be created by thermal freeze-out. If $\kappa$ is sufficiently small, the direct detection limit can be bypassed and the dark matter relic density generated via freeze-in as mentioned in in Section \ref{sec:freeze-in}.

This ends our discussion of the pure singlet case. For the remainder of this paper, we focus on the triplet, quintuplet and quadruplet models.

\begin{figure}[th]
  \begin{subfigure}{0.5\textwidth}
    \centering
    \includegraphics[width=\linewidth]{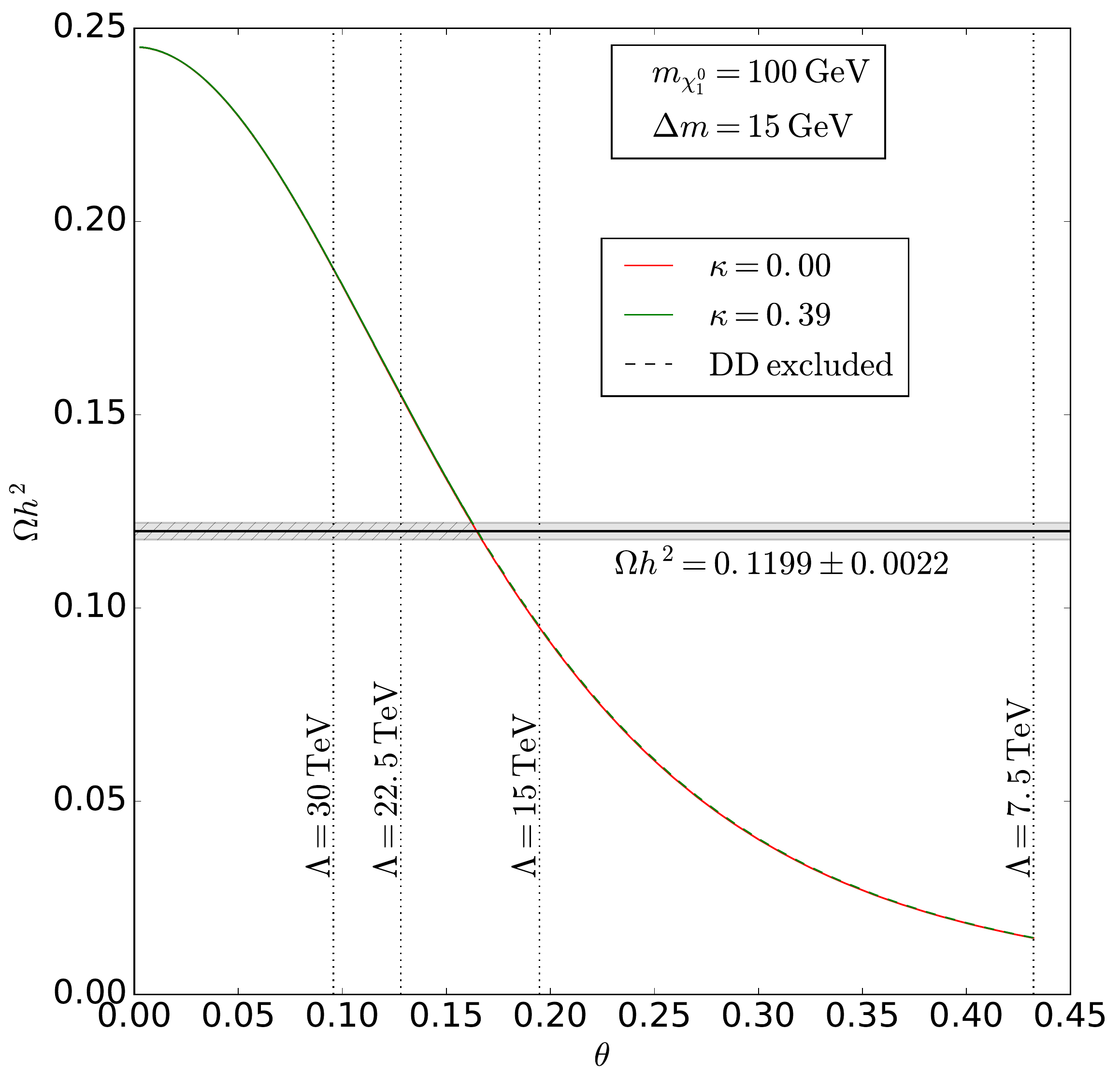}
    \caption{}
  \end{subfigure}
  \begin{subfigure}{0.5\textwidth}
    \centering
    \includegraphics[width=\linewidth]{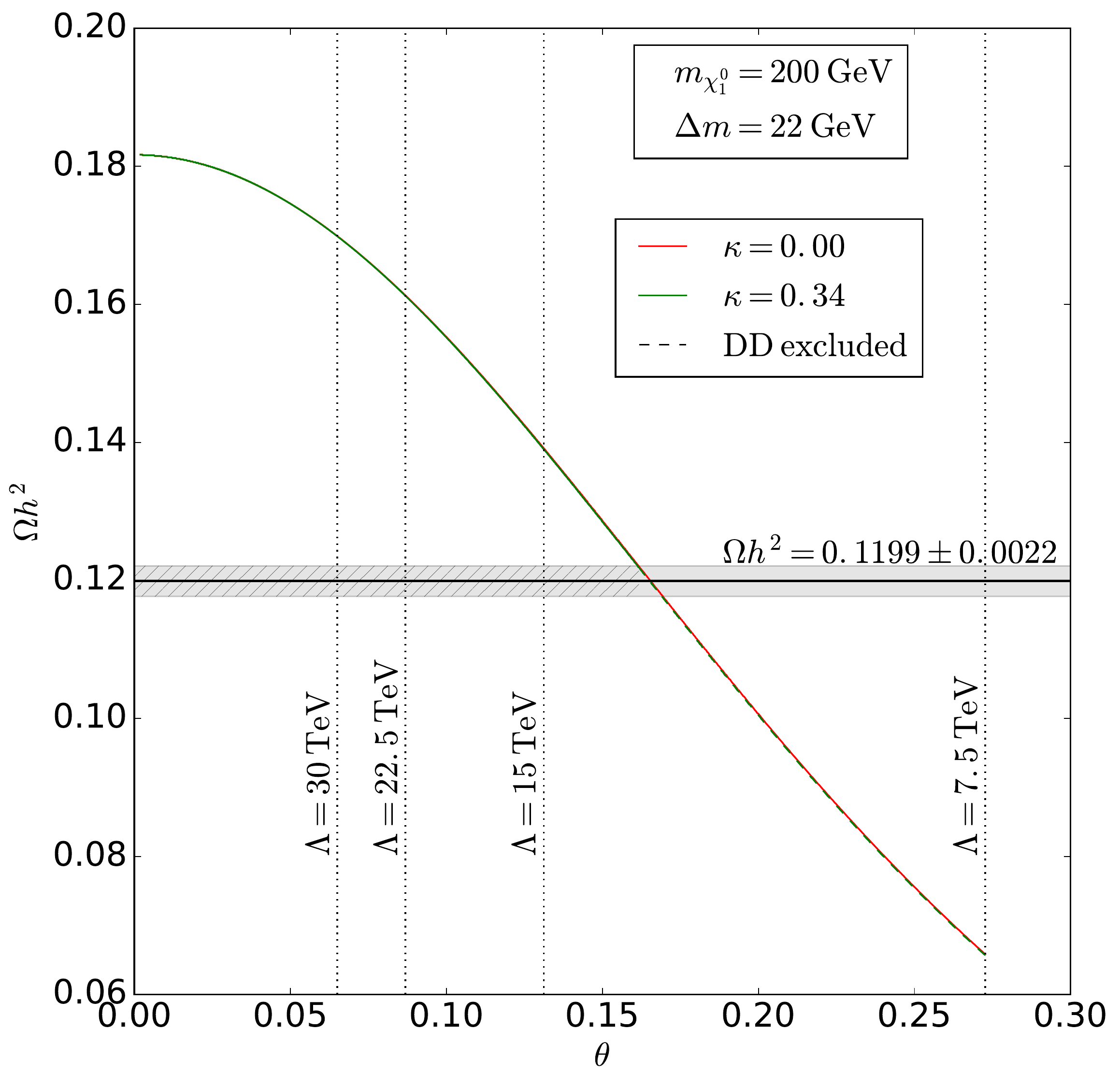}
    \caption{}
  \end{subfigure}
  \caption{Thermal relic density as a function of the singlet-triplet mixing angle $\theta$ for a WIMP mass of $m_{\chi^0_1}=100$ GeV (left panel) and $m_{\chi^0_1}=200$ GeV (right panel). For a fixed Wilson coefficient $\lambda=1$ we also indicate the suppression scale $\Lambda$ corresponding to the respective mixing angles assuming Eq.~\eqref{eq:theta3}. The impact of ${\cal L}_{\rm quartic}$ is negligible, unless the Wilson coefficient $\kappa$ is so large that the parameter point is already excluded by direct detection.}
  \label{TSM_Omega}
\end{figure}

\begin{figure}[th]
  \begin{subfigure}{0.5\textwidth}
    \centering
    \includegraphics[width=\linewidth]{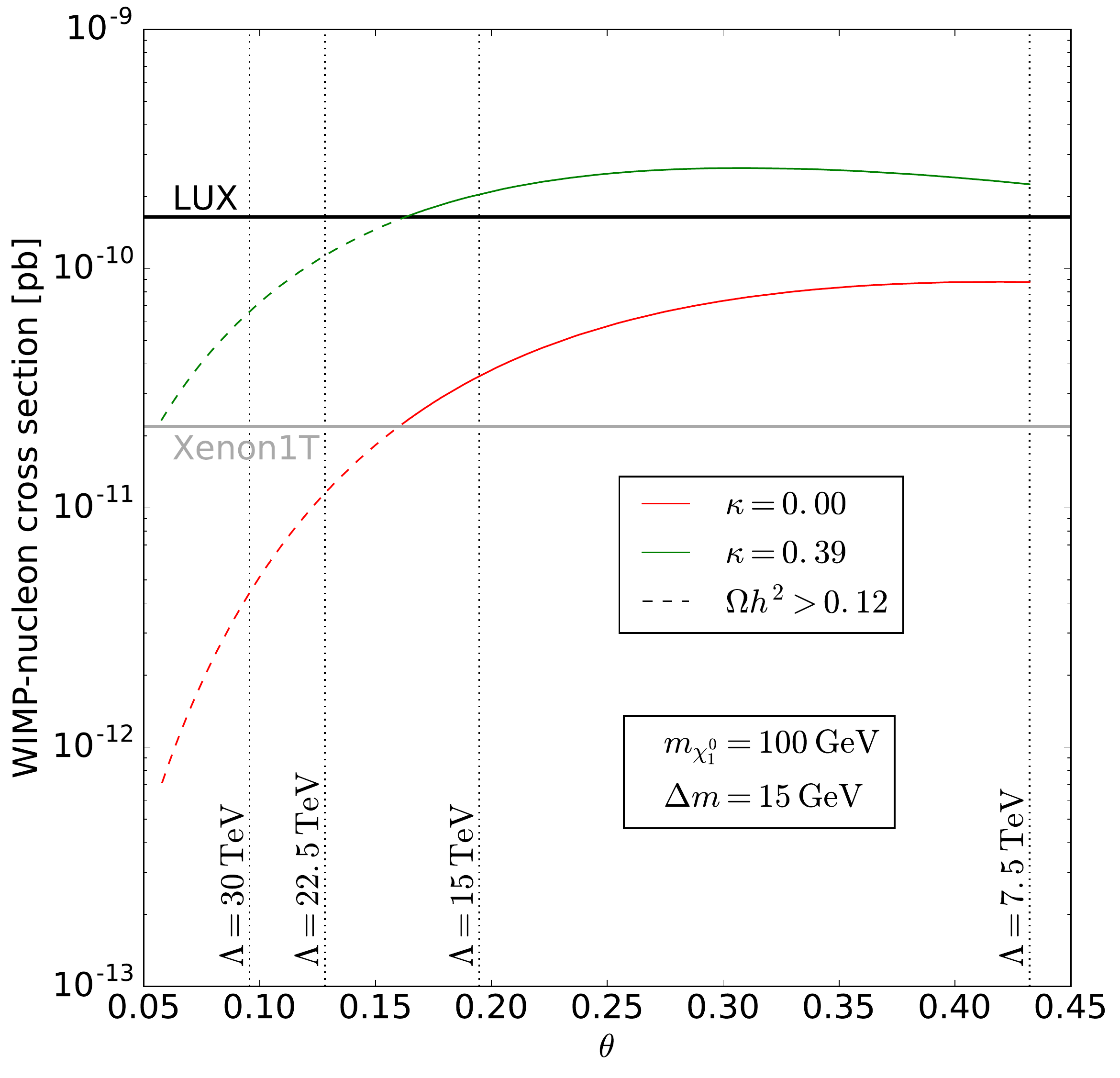}
    \caption{}\label{TSM_DD-a}
  \end{subfigure}
  \begin{subfigure}{0.5\textwidth}
    \centering
    \includegraphics[width=\linewidth]{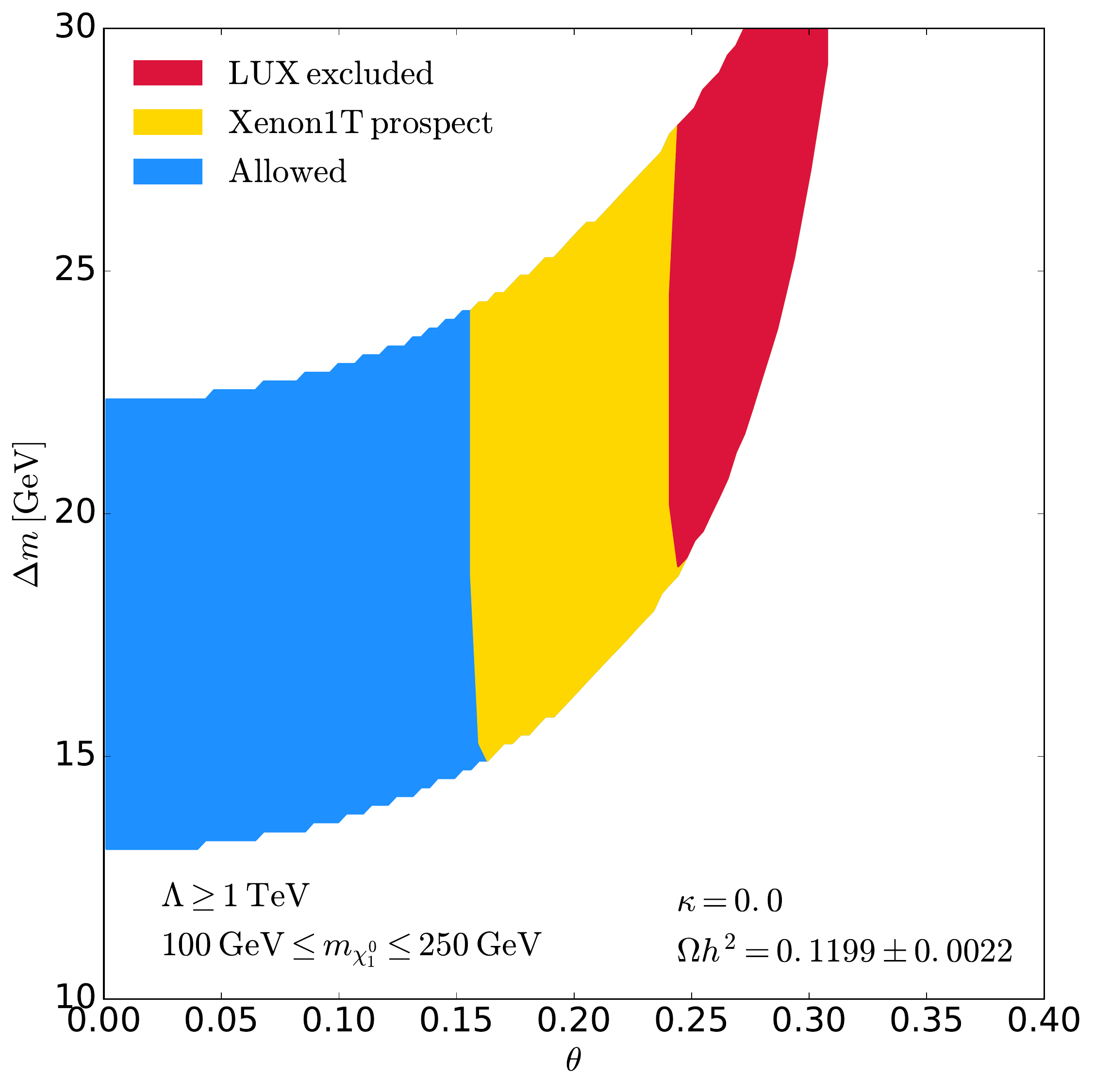}
    \caption{}
  \end{subfigure}
  \caption{Left panel: WIMP-nucleon cross section in the triplet model for $m_{\chi^0_1}=100$ GeV and a mass splitting of $\Delta m = 15$ GeV. Here $\theta$ is the triplet-singlet mixing angle. Direct detection bounds require the Wilson coefficient $\kappa$ to be rather small, $\kappa\lesssim 0.39$.} Right panel: Here $\kappa=0$ and the observed thermal relic density has been imposed as a constraint; in the coloured regions there exists a solution for some $\Lambda$ and $m_{\chi^0_1}$. 
  Even for $\kappa=0$ direct detection bounds are becoming important.
  \label{TSM_DD}
\end{figure}

The results for the relic density and spin-independent WIMP-nucleon cross section for some representative parameter points in the triplet model are shown in Figs.~\ref{TSM_Omega} and \ref{TSM_DD} respectively. To be specific, Fig.~\ref{TSM_Omega} shows the relic density as a function of the mixing angle for two fixed values of the WIMP mass and the mass splitting $\Delta m=m_{\chi^0_2}-m_{\chi^0_1}=\xi$, where $\xi$ is defined in Eq.~\eqref{eq:delm-trip}. For a fixed value of $\lambda$, and assuming Eq.~\eqref{eq:theta3} holds, the mixing angle is in one-to-one correspondence with the suppression scale $\Lambda$, which we also provide in the plots. For definiteness we set $\lambda=1$ for the Wilson coefficient governing the mixing.\footnote{Of course this does not necessarily imply that the states inducing the mixing operator have ${\cal O}(1)$ couplings and ${\cal O}(\Lambda)$ masses --- by a simultaneous rescaling of $\lambda$ and $\Lambda$ the same mixing operator could result from more weakly coupled states with correspondingly lower masses.}

Fixing the parameters $\theta$, $m_{\chi^0_1}$ and $\Delta m$ is not enough to completely determine the thermal relic density, since the annihilation cross section also depends on $\kappa$. The (co-)annihilation relies on both mixing of the singlet and the triplet which is determined by $\theta$, as well as the quartic coupling of the singlet to the Higgs boson $\kappa$. We see that as $\theta$ increase the (co-) annihilation increases and $\Omega h^2$ decreases, however this is independent of any of the values of $\kappa$ shown on the plot.

The chosen values of $\kappa$ can be understood from Fig.~\ref{TSM_DD-a} where we show $\sigma_{\rm SI}$ as a function of $\theta$. We see that as soon as $\kappa$ reaches values around $0.4$ (in units where $\lambda=1$) the direct detection limits become relevant, at least for mixing angles giving the observed relic density.
Any value of $\kappa$ large enough for the relic density to depart significantly from the $\kappa=0$ prediction (indicated by the red line in Fig.~\ref{TSM_Omega}) will lead to a direct detection cross section which is too large to be compatible with LUX. For example, mixing angles below about $0.16$ in Figs.~\ref{TSM_Omega} and \ref{TSM_DD-a} cannot give the observed relic density of $\Omega h^2=0.1199\pm0.0022$~\cite{planck}, since this would require a too large value of $\kappa$, as indicated by the hatched area. (However, this limiting value of the $\theta$ of course depends on the choice of $m_{\chi^0_1}$ and of $\Delta m$.) 
The reason that $\Omega h^2$ depends less strongly on $\kappa$ than $\sigma_{\rm SI}$ is that the contribution of coannihilation for the relic density means that diagrams mediated by a Higgs are less significant.
In Fig.~\ref{TSM_DD}, we also show that the future Xenon1T experiment has the potential to severely constrain the parameter space of our model even at $\kappa=0$.

\begin{figure}[th]
  \begin{subfigure}{0.5\textwidth}
    \centering
    \includegraphics[width=\linewidth]{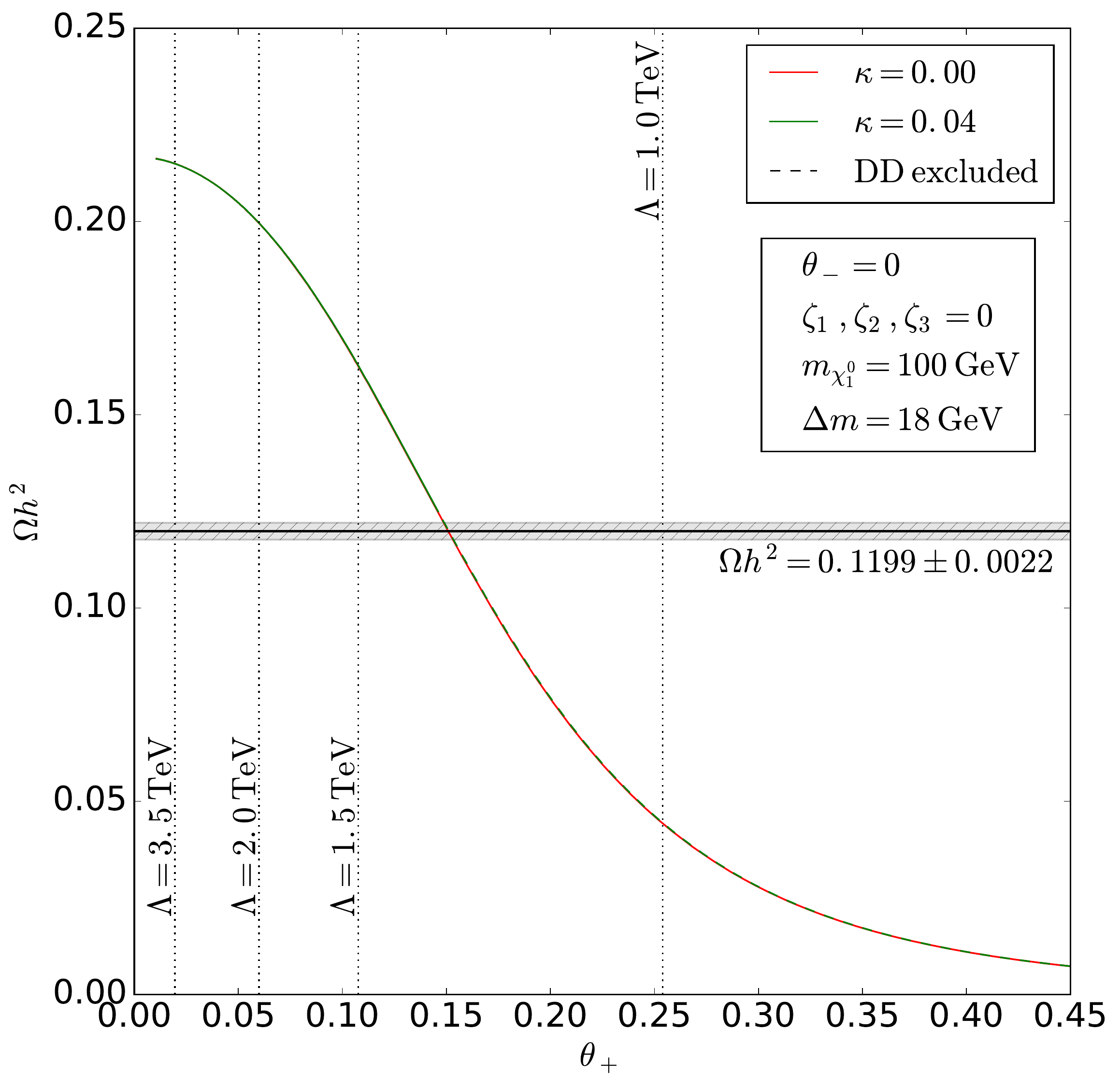}
    \caption{}\label{QQSM_Omega-a}
  \end{subfigure}
  \begin{subfigure}{0.5\textwidth}
    \centering
    \includegraphics[width=\linewidth]{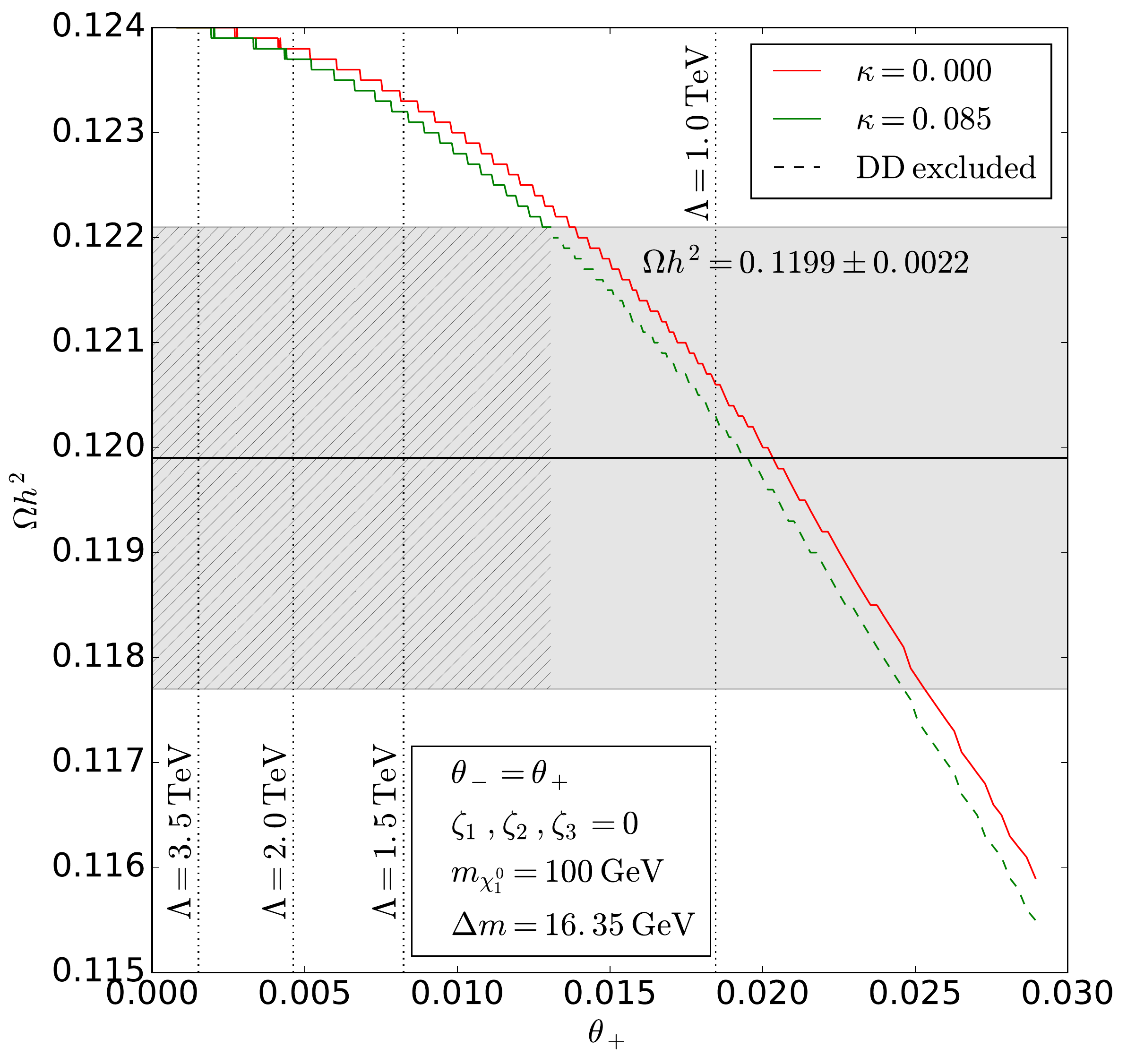}
    \caption{}\label{QQSM_Omega-b}
  \end{subfigure}
  \caption{The relic density as a function of the mixing angle $\theta_+$ in the quadruplet model for $m=100$ GeV. Left panel: $\theta_-=0$, representing the generic case of one dominant mixing angle. Right panel: the fine-tuned case $\theta_+=\theta_-$.}
  \label{QQSM_Omega}
\end{figure}

The corresponding results for $\Omega h^2$ in the quadruplet model are shown in Figs.~\ref{QQSM_Omega} for a WIMP mass of 100 GeV and two particular choices for the mixing angle $\theta_-$. Precise definitions of the mixing angles and the physical masses in terms of the parameters of the Lagrangian can be found in App.~\ref{sec:quad}. In Fig.~\ref{QQSM_Omega-a} we show the case $\theta_-=0$, in order to represent the generic case of one mixing angle being dominant and the other having a small impact. In Fig.~\ref{QQSM_Omega-b} we have set $\theta_-=\theta_+$. This latter case is rather fine-tuned, since it requires that the Wilson coefficients be chosen precisely such that one compensates for the relative suppression of the mixing angles by the factor $(M-m)/(M+m)$, see Eq.~\eqref{eq:theta-quad}. The fine-tuning is reflected in the numerical artifacts appearing in the curves, and is also apparent from the fact that it requires a very precise choice of $\Delta m$ in order to obtain the correct relic density. We clearly see that $\theta_+$ is suppressed on comparing Fig.~\ref{QQSM_Omega-b} to Fig.~\ref{QQSM_Omega-a}. In these plots we have set $\kappa'=\zeta_{1,2,3}=0$. We have verified that the impact of nonzero $\kappa'$ is small (but noticeable) after absorbing the mass shift it induces into the definition of the mass parameter $M$, since its effect on the annihilation cross section is suppressed by the mixing angle, details can be found in App.~B. By contrast, the relic density is rather sensitive to the values of $\zeta_{1,2,3}$, since a nonzero $\zeta_{1,2,3}$ can easily shift the mass differences between the charged and neutral $n$-plet-like states by several GeV, which will drastically affect the coannihilation rate.

\begin{figure}[th]
  \begin{subfigure}{0.5\textwidth}
    \centering
    \includegraphics[width=\linewidth]{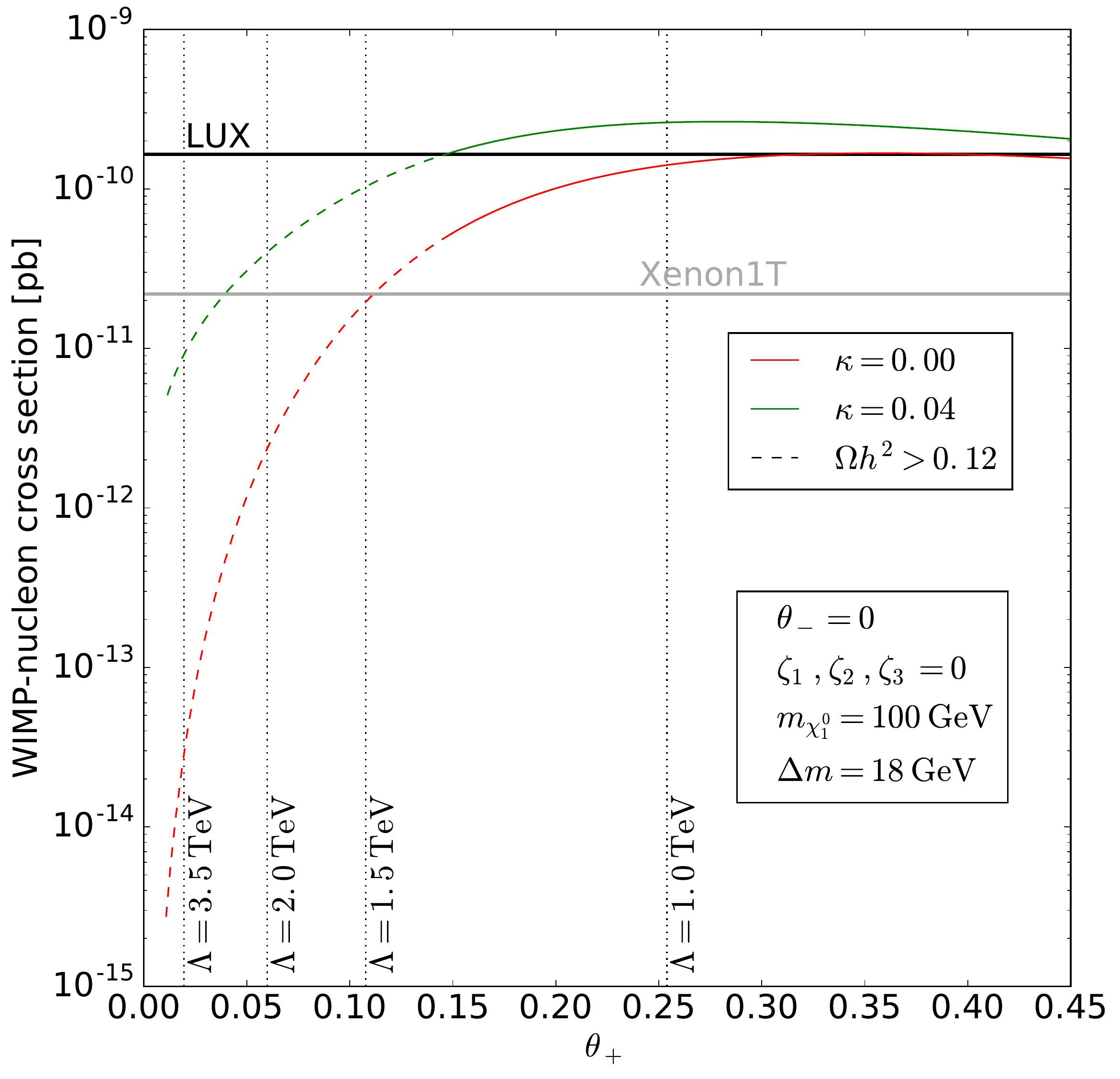}
    \caption{}\label{QQSM_DD-a}
  \end{subfigure}
  \begin{subfigure}{0.5\textwidth}
    \centering
    \includegraphics[width=\linewidth]{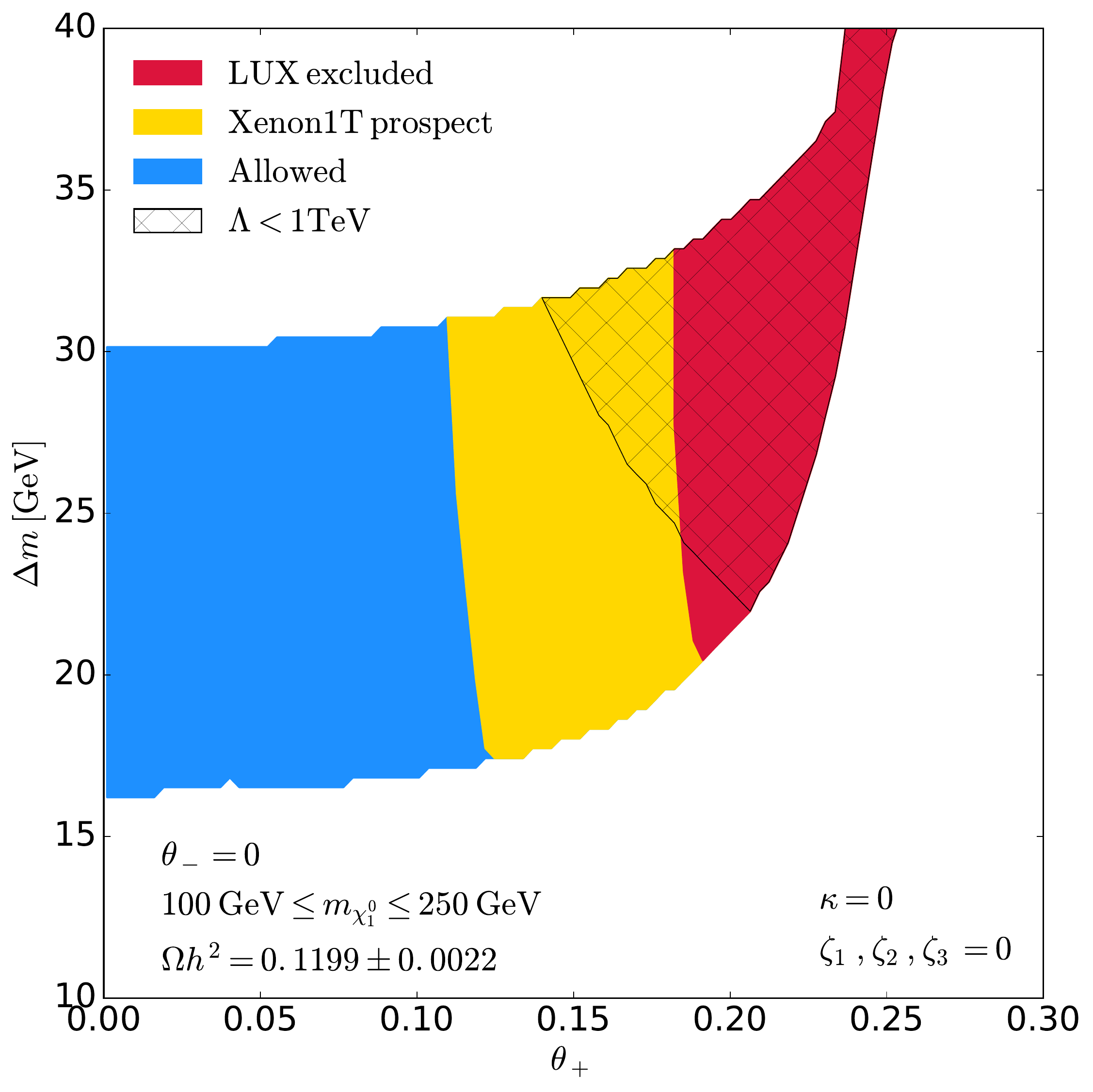}
    \caption{}\label{QQSM_DD-b}
  \end{subfigure}
  \caption{Same as Fig.~\ref{TSM_Omega} but for the quadruplet model at $\theta_-=0$. Left panel: the model is excluded by direct detection for all but small $\kappa\lesssim 0.04$. Right panel: Current and projected exclusion bounds at $\kappa=0$. The hatched region corresponds to large mixing angles requiring a scale $\Lambda<1$ TeV, where our effective theory can no longer be trusted.}
  \label{QQSM_DD}
\end{figure}

Figs.~\ref{QQSM_DD} show the direct detection bounds on the quadruplet model with  $\theta_-=0$. We have set $\kappa'=\zeta_{1,2,3}=0$ for concreteness, after verifying that their impact on direct detection is in any case subdominant. In Fig.~\ref{QQSM_DD-a} we plot the spin-independent cross section as a function of the mixing angle for a fixed $m=100$ GeV; once more, direct detection constrains the quartic coupling $\kappa$ to be tiny. Fig.~\ref{QQSM_DD-b} shows the parameter space for $\kappa=0$, which is now starting to be constrained by direct detection bounds. However, when demanding a cutoff scale $\Lambda>1$ TeV (which is reasonable to ensure the validity of our effective theory), the bounds from LUX are not restricting this particular slice of parameter space yet. One further observes that the relation of $\theta_+$ to $\Lambda$ in these plots differ from that in Fig.~\ref{TSM_DD}, due to the $v/\Lambda$ suppression factor in the mixing angle which arises on increasing $n$, therefore lower values of $\Lambda$ can provide the observed relic density in regions of parameter space which can be probed by direct detection experiments.

\begin{figure}[th]
  \begin{subfigure}{0.5\textwidth}
    \centering
    \includegraphics[width=\linewidth]{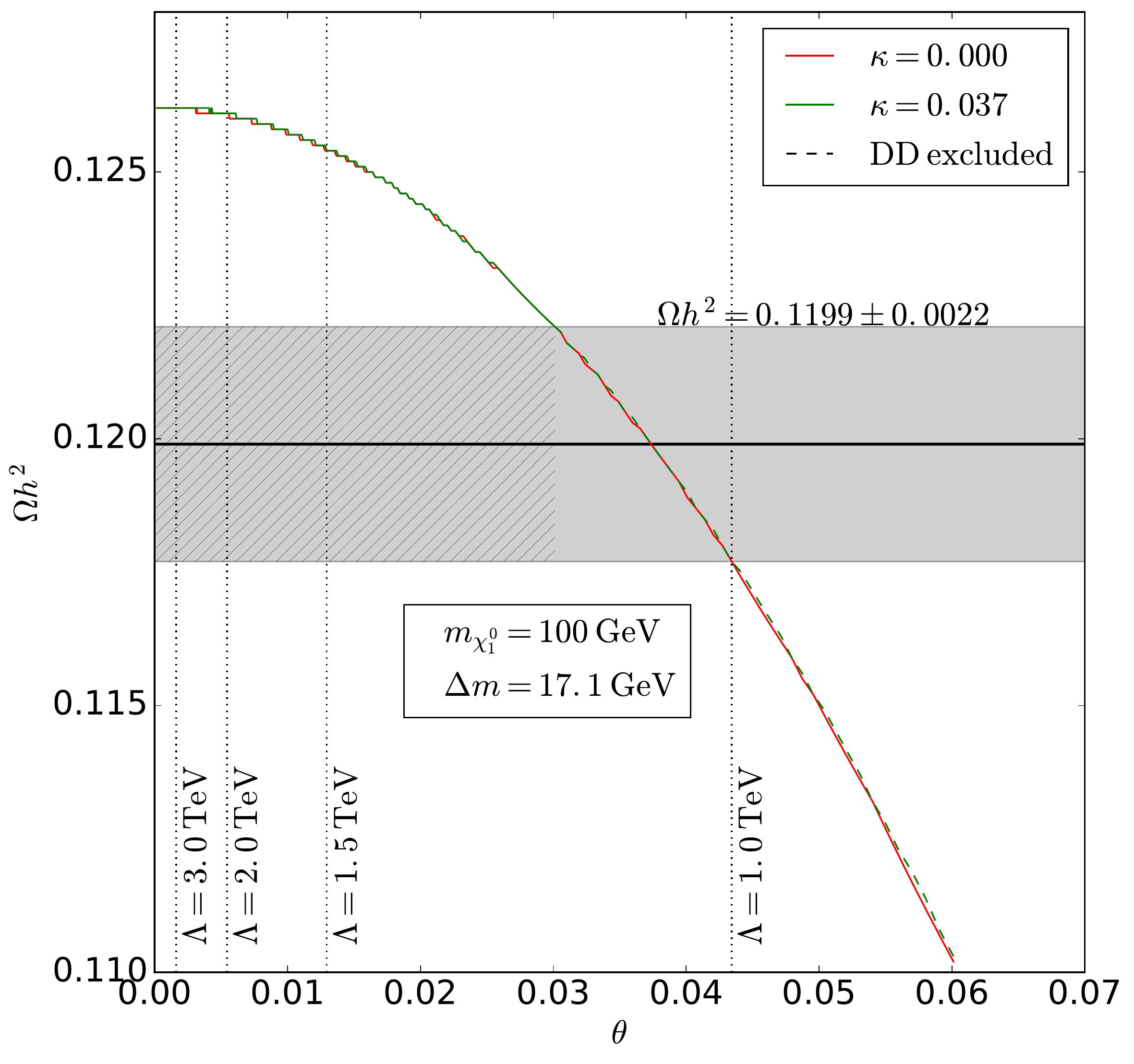}
    \caption{}
  \end{subfigure}
  \begin{subfigure}{0.5\textwidth}
    \centering
    \includegraphics[width=\linewidth]{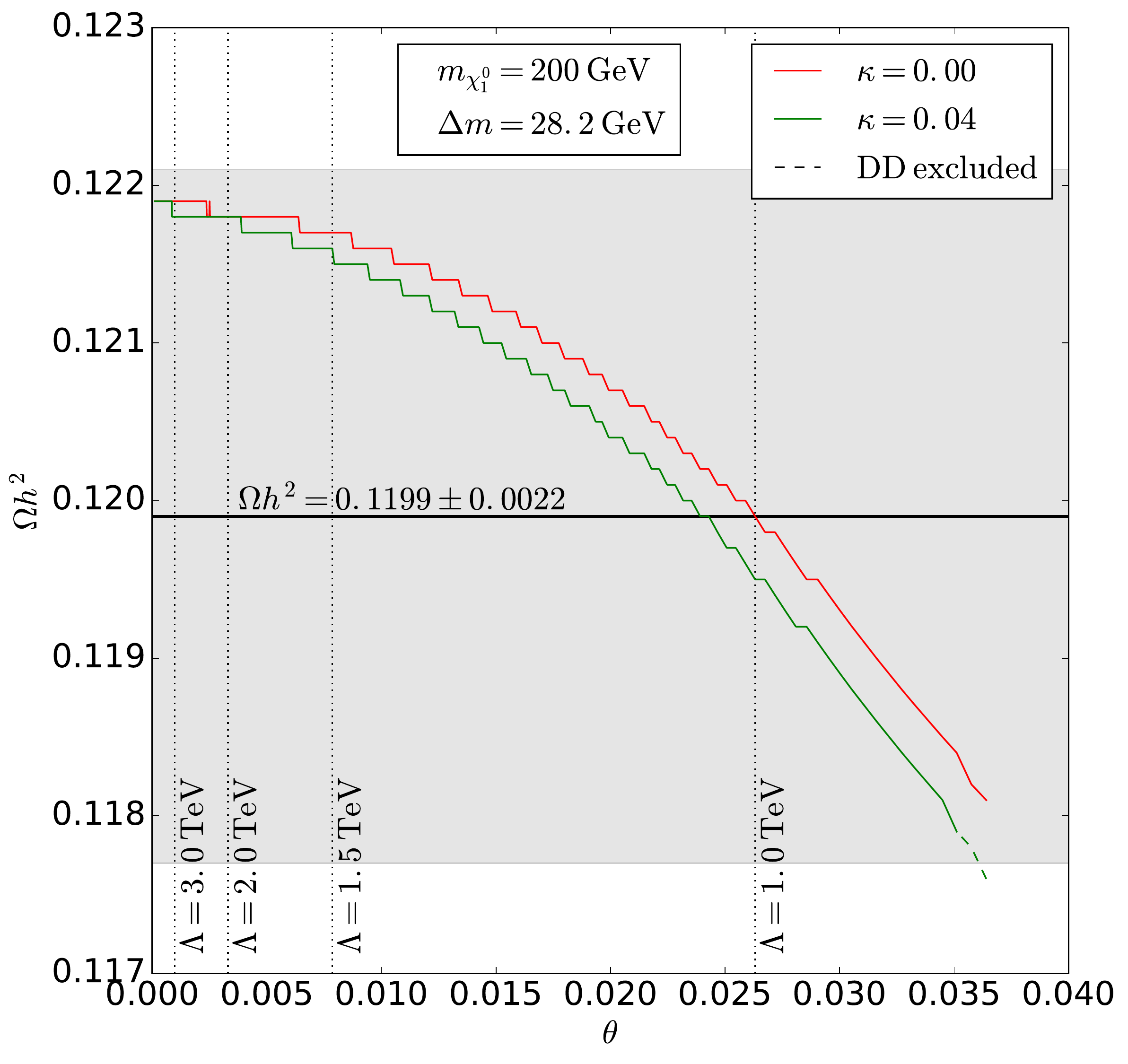}
    \caption{}
  \end{subfigure}
  \caption{Same as Fig.~\ref{TSM_DD} but for the quintuplet model, with appropriately modified mass splittings $\Delta m$.}
  \label{QSM_Omega}
\end{figure}

\begin{figure}[th]
  \begin{subfigure}{0.5\textwidth}
    \centering
    \includegraphics[width=\linewidth]{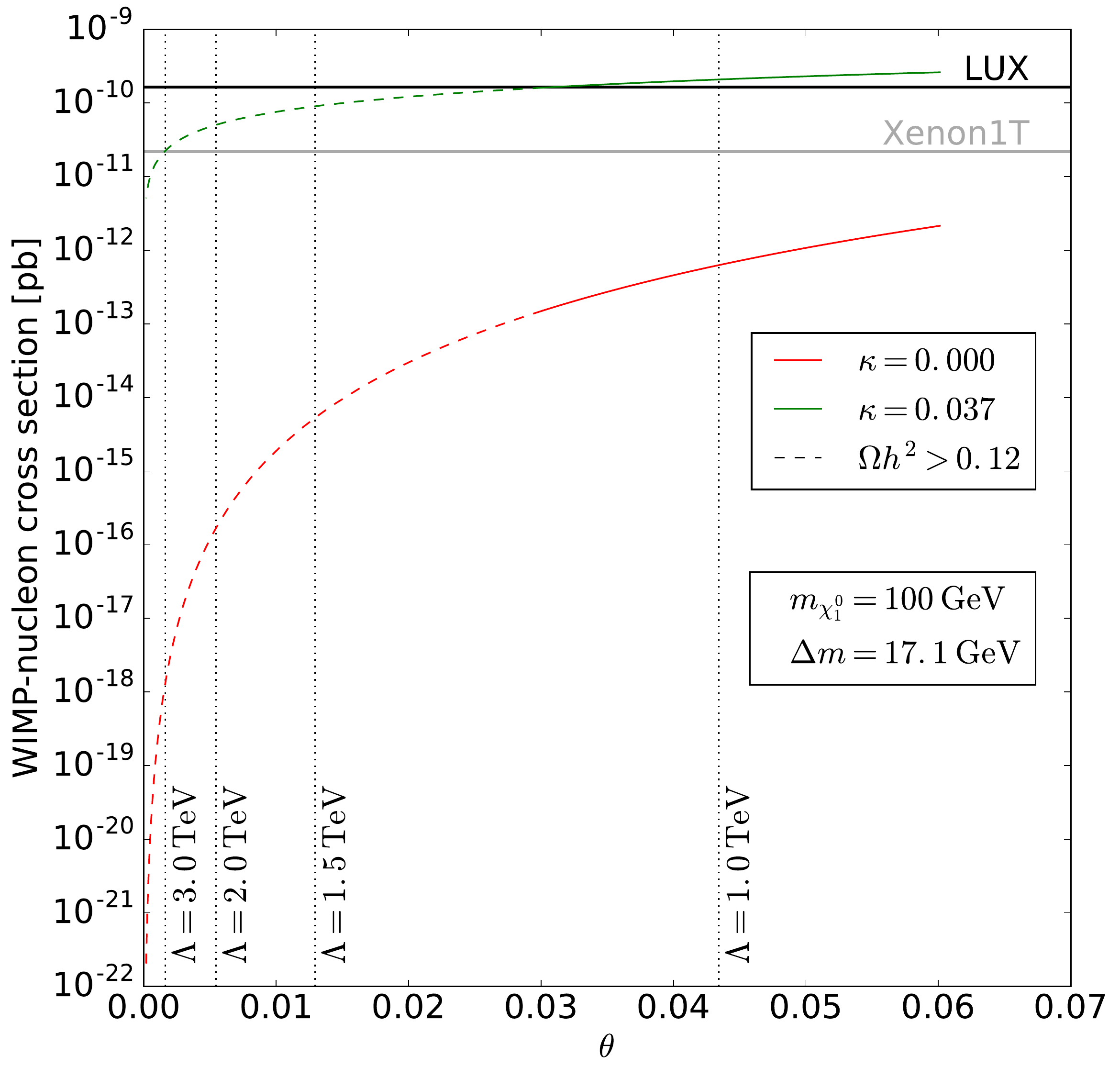}
    \caption{}
  \end{subfigure}
  \begin{subfigure}{0.5\textwidth}
    \centering
    \includegraphics[width=\linewidth]{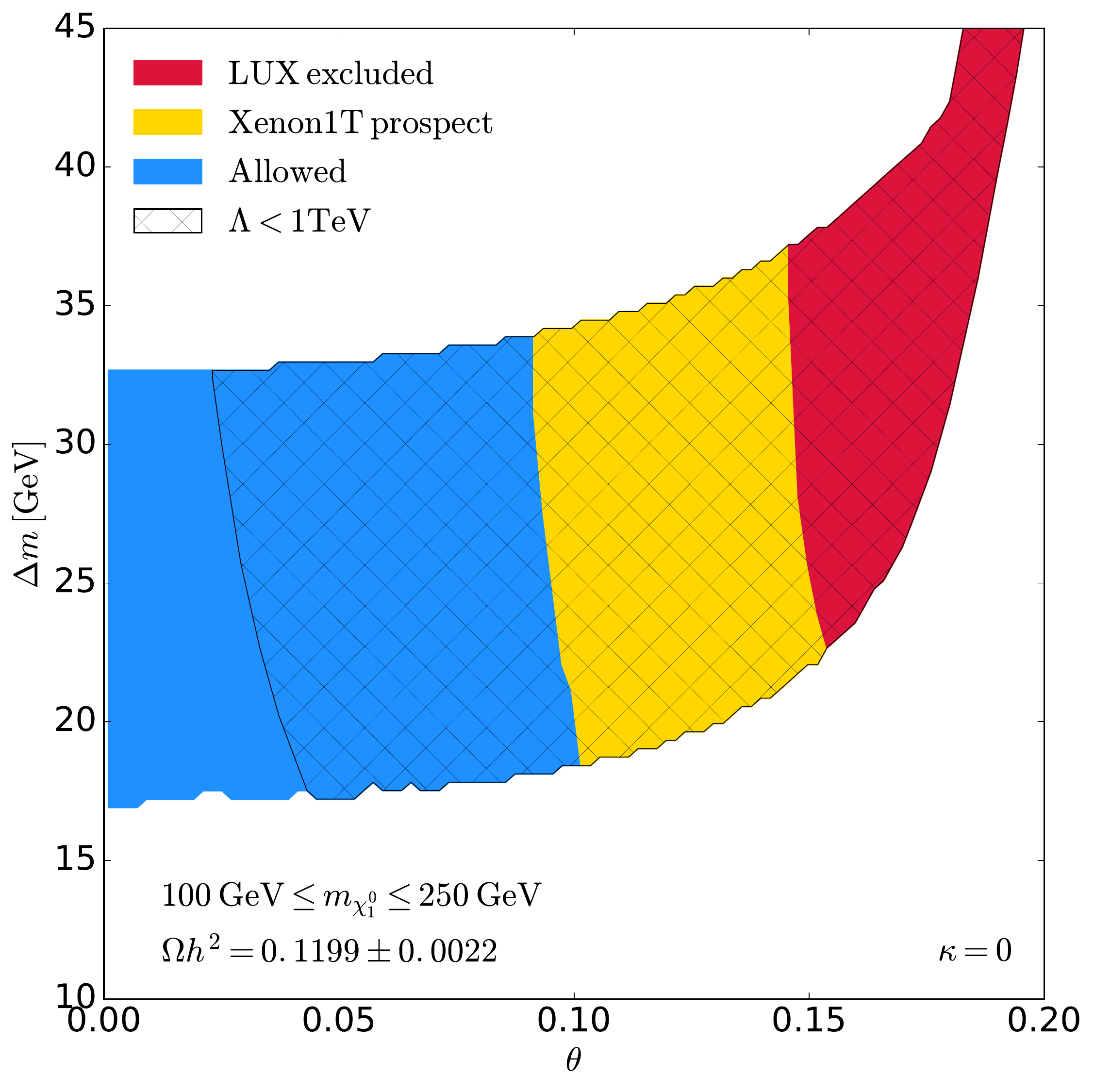}
    \caption{}\label{QSM_DD-b}
  \end{subfigure}
  \caption{Same as Fig.~\ref{TSM_Omega} but for the quintuplet model. Left panel: the model is excluded by direct detection for all but small $\kappa$. Right panel: Current and projected exclusion bounds at $\kappa=0$.}
  \label{QSM_DD}
\end{figure}

Finally, for the quintuplet model Figs.~\ref{QSM_Omega} and \ref{QSM_DD} are equivalent to Figs.~\ref{TSM_Omega} and \ref{TSM_DD} respectively. In the quintuplet model the mixing angle arises from a dimension-7 operator, so when demanding $\Lambda>1$ TeV the mixing angle is generically very small. This suppression is of the order $v^2/\Lambda^2$ compared to the triplet case, as can be seen in Eq.~(\ref{eq:theta-quint}) and further details can be found in App.~C. Consequently, the difference between the parameters $M-m$ needs to be quite finely adjusted in order to reproduce the observed relic density. Nevertheless a suitable choice can always be found. While the dependence on $\theta$ is qualitatively similar to the triplet and quadruplet case, there is one interesting distinguishing feature: the influence of $\kappa$ on the relic density is significantly more pronounced in the quintuplet model. This is because the mixing operator, being dimension-7, is now relatively less important than ${\cal L}_{\rm quartic}$ which remains dimension-5. However, the direct detection bounds are still too stringent to allow for a significant change in the relic density due to a nonzero $\kappa$. The green curves in Fig.~\ref{QSM_Omega} correspond to the maximal value which $\kappa$ can take without violating the LUX bound; evidently the predicted relic density does not change much compared to the $\kappa=0$ case, illustrated by the red curves.
Moreover, in Fig.~\ref{QSM_DD-b} we observe that, due to the relative suppression of the mixing angle in the quintuplet case, the entire region with the correct relic density that has been probed by LUX and that is expected to be probed by Xenon1T lies below the limit $\Lambda<1$ TeV. 

Fig.~\ref{QSM_DD-b} shows that at $\kappa=0$, the mixing angles induced by the dimension-7 operator ${\cal L}_{\rm mix}$ are generically too small to be relevant for direct detection as of now (unless one chooses a suppression scale $\Lambda<1$ TeV, for which our effective theory becomes unreliable).

In summary, we find that in all our models the relic density must be dominated by $n$-plet-singlet mixing effects rather than by annihilation through ${\cal L}_{\rm quartic}$. In the case of the triplet model, where both ${\cal L}_{\rm quartic}$ and ${\cal L}_{\rm mix}$ are dimension-5, and if the Wilson coefficients $\kappa$ and $\lambda$ are of similar magnitude, this is partly a consequence of dark matter annihilation through ${\cal L}_{\rm quartic}$ being $p$-wave suppressed. However, even more important is the direct detection constraint, which forces our models into the parameter space region of small $\kappa\ll\lambda$. Possible explanations for this will obviously depend on the UV completion, and we will not investigate them here. However, we note that when UV-completing the triplet-singlet model by the MSSM, one indeed finds a numerically somewhat suppressed value of $\kappa/\lambda=\tan\theta_w\approx 0.53$, see App.~A.

We have further investigated the indirect detection prospects of our models, again by using {\tt micrOMEGAs 4.3.1} to calculate the present-day annihilation cross section, which we find to be predominantly to $W^+W^-$. We find that, for the region of parameter space respecting the relic density and direct detection constraints, the annihilation rate is well below the current bounds~\cite{Ahnen:2016qkx}. This is due to the fact that the relic density is mainly generated via coannihilation channels, such that the self-annihilation rates are low. This would not necessarily be the case for heavier dark matter candidates where the Sommerfeld effect would affect the result, for bino-wino mixing in the MSSM see~\cite{Beneke:2016ync,Beneke:2016jpw}, but for those masses of interest to the LHC this is not relevant.

\section{Conclusions and outlook}
\label{sec:conclusions}

We have studied simple effective models for WIMP dark matter, where the electroweak-scale particle content is given by a fermionic singlet and a fermionic $\SU{2}$ $n$-plet with $n\geq 3$ in addition to that of the Standard Model,. The neutral component of the $n$-plet and the singlet mix through a higher-dimensional operator. The resulting state is WIMP dark matter candidate, whose relic abundance is obtained from freeze-out with the assistance of coannihilation channels.

These models should be thought of as allowing an effective description of the dark matter and collider properties of some more complete model. The corresponding scale at which this description breaks down could be as low as a TeV, but any further states appearing at such a scale would generically be too heavy to be discovered at the LHC if they only carry electroweak charges.

We concentrated on the parameter space of the models for $n=3,4,5$, which is quite constrained by direct detection experiments. In particular, direct singlet annihilation into a pair of Higgs bosons through a Higgs portal-like dimension-5 operator ${\cal L}_{\rm quartic}$ must be subdominant in order for the model to be viable. Therefore, as far as the model is compatible with direct detection, its relic density is essentially dictated by the singlet-$n$-plet mixing angle.  Note that while for $n=3$ the effect of the constrained quartic coupling $\kappa$ is negligible, for larger $n$ the mixing is dimensionally suppressed compared to the quartic coupling, and therefore the effect of $\kappa$ becomes somewhat more pronounced. The requirement that the Wilson coefficient of ${\cal L}_{\rm quartic}$ be small translates into a nontrivial constraint on possible UV completions. We moreover find that the next generation of direct detection experiments will probe the model even further even if such a coupling is negligibly small.  

For a given DM mass and splitting between the singlet and $n$-plet, the relic density constraint generically fixes the mixing angle, which translates into a certain effective theory cut-off scale $\Lambda$. As expected, we see that for larger mixings the mass splittings must be larger as the principal mechanism dictating the relic density is coannihilation. We further mention that the three cases $n=3,4$ and 5 differ due to the relative suppression of the mixing angle by factors of $v/\Lambda$ compared to the triplet case, meaning that as $n$ increases smaller scales $\Lambda$ are probed via direct detection experiments.  However, at very small mixing angles the dependence of the relic density becomes effectively $\theta$-independent, so $\Lambda$ is not bounded from above (except eventually by the requirement that the singlet-like WIMP should be in thermal equilibrium with the $n$-plet-like states). 

We have further investigated the indirect detection prospects of this model, finding these not be constraining for the parameter space of interest to us.

The region we have chosen to study corresponds to electroweak-scale WIMP masses, since this region is kinematically accessible at the LHC. In fact, collider searches for supersymmetric neutralinos and charginos will constrain the allowed parameter space even further. Relevant analyses could be searches for (ISR) jets and missing energy, possibly including leptons if the singlet-$n$-plet mass splitting is large enough, disappearing track searches \cite{Aad:2013yna, CMS:2014gxa}, displaced vertex searches to constrain  $\chi^0_2\into\chi^0_1$ decays \cite{Nagata:2015pra}, or searches for signatures specific to $\chi^{++}$ decay.
We will address the collider phenomenology of the models we have presented in a forthcoming publication \cite{BBDR}.

\subsection*{Acknowledgements}

The authors thank G.~B\'elanger and A.~Pukhov for correspondence. This work was supported by the LabEx ``Origins, Constituents and Evolution of the Universe'' (ANR-11-LABX-0060) and the
A*MIDEX project (ANR-11-IDEX-0001-02) funded by the ``Investissements d'Avenir'' French government programme,
managed by the ANR. 

\appendix

\section*{Appendix A: The triplet-singlet model}
\label{sec:trip}
This appendix contains some technical details on how to obtain the effective wino-bino Lagrangian from the full electroweak sector of the MSSM, and how to generalize this result to the non-supersymmetric case. We keep the discussion rather detailed in the interest of clarity, although the results are well known (see e.g.~\cite{Nagata:2015pra}).

In the well-tempered bino-wino scenario in the MSSM, dark matter is constituted of a Majorana singlet $\wt B$ and a Majorana triplet $\wt W^a$. These couple to the Higgsinos $\tilde h_u$ and $\tilde h_d$ via the Lagrangian\footnote{We are ignoring the gauge boson couplings, which can be restored by replacing all derivatives by covariant derivatives.} (in two-component Weyl spinor notation)
\be\begin{split}
{\cal L}=\;&\tilde h_u^\dag i\ol\sigma^\mu\partial_\mu\tilde h_u+\tilde h_d^\dag i\ol\sigma^\mu\partial_\mu\tilde h_d+\wt W^{a\dag} i\ol\sigma^\mu\partial_\mu\wt W^a+\wt B^\dag i\ol\sigma^\mu\partial_\mu\wt B\\
&-\frac{1}{2} M_1\,\wt B\wt B-\frac{1}{2} M_2\,\wt W^a\wt W^a-\mu\,\tilde h_u\tilde h_d\hc\\
&-\sqrt{2}g\left(h_u^\dag \tau^a\tilde h_u\right)\wt W^a-\frac{g'}{\sqrt{2}}\,h_u^\dag \tilde h_u \wt B-\sqrt{2}g\left(h_d^\dag \tau^a\tilde h_d\right)\wt W^a-\frac{g'}{\sqrt{2}}h_d^\dag \tilde h_d\wt B\hc
\end{split}
\ee
where $\tau^a=\sigma^a/2$ generate the fundamental representation of $\SU{2}$. We are taking $M_1$, $M_2$, $\mu$ real and positive for simplicity. The case of interest is $\mu\gg M_1, M_2, m_Z$. The equations of motion for $\tilde h_u$ and $\tilde h_d$ read, at the zero-derivative level,
\be\begin{split}
\tilde h_u=\frac{1}{\mu}\left(-\sqrt{2}g\,h_d^\dag \tau^a\wt W^a-\frac{g'}{\sqrt{2}}\,h_d^\dag \wt B\right)\,,\\
\tilde h_d=\frac{1}{\mu}\left(-\sqrt{2}g\,h_u^\dag \tau^a\wt W^a+\frac{g'}{\sqrt{2}}\,h_u^\dag \wt B\right)\,.
\end{split}
\ee
Substituting back into ${\cal L}$ one obtains the leading-order effective Lagrangian for the light degrees of freedom $\wt B$ and $\wt W^a$,
\be\begin{split}\label{eq:LeffBW}
{\cal L}_{\rm eff}=\;&\wt W^{a\dag} i\ol\sigma^\mu\partial_\mu\wt W^a+\wt B^\dag i\ol\sigma^\mu\partial_\mu\wt B\\
+&\frac{1}{\mu^2}\left(\sqrt{2}g\,\wt W^{a\dag} \tau^a h_d+\frac{g'}{\sqrt{2}}\,\wt B^\dag h_d\right)i\ol\sigma^\mu\partial_\mu\left(\sqrt{2}g\,h_d^\dag \tau^a\wt W^a+\frac{g'}{\sqrt{2}}\,h_d^\dag \wt B\right)\\
+&\frac{1}{\mu^2}\left(\sqrt{2}g\,\wt W^{a\dag} \tau^a h_u-\frac{g'}{\sqrt{2}}\,\wt B^\dag h_u\right)i\ol\sigma^\mu\partial_\mu\left(\sqrt{2}g\,h_u^\dag \tau^a\wt W^a-\frac{g'}{\sqrt{2}}\,h_u^\dag \wt B\right)\\
-&\left(\frac{1}{2} M_1\,\wt B\wt B+\frac{1}{2} M_2\,\wt W^a\wt W^a\hc\right)\\
+&\frac{1}{\mu}\left(\sqrt{2}g\,h_u^\dag \tau^a\wt W^a+\frac{g'}{\sqrt{2}}\,h_u^\dag \wt B\right)\left(\sqrt{2}g\,h_d^\dag \tau^a\wt W^a-\frac{g'}{\sqrt{2}}\,h_d^\dag \wt B\right)\hc\\
   \end{split}
\ee
The dimension-6 terms are corrections for the wino, bino and Higgs kinetic terms. To compare with the known chargino and neutralino mass matrices in the literature, we set $h_u={0\choose v\sb}$ and $h_d={{v\cb}\choose 0}$ as well as $g'\frac{v}{\sqrt{2}}=s_w m_Z$ and $g\frac{v}{\sqrt{2}}=c_w m_Z$. In terms of $\psi_0={\wt B\choose\wt W^3}$ we obtain the effective neutralino Lagrangian
\be
\begin{split}
 {\cal L}_{\rm eff, neut}=\psi_0^\dag \,G_0\,i\ol\sigma^\mu\partial_\mu\psi_0-\frac{1}{2}\psi_0^T \wh M_0 \psi_0\hc
\end{split}
\ee
where the kinetic and mass matrices are given, up to terms suppressed by $1/\mu^3$, by
\be
G_0=\mathbbm{1}+\frac{m_Z^2}{\mu^2}\;T\,,\qquad \wh M_0=\left(\begin{array}{cc} M_1 & \\ & M_2\end{array} \right)-\stb\frac{m_Z^2}{\mu}\;T
\ee
with
\be
T=\left(\begin{array}{cc}s_w^2 & -s_w c_w \\ -s_w c_w & c_w^2\end{array}\right)\,.
\ee
Defining 
\be
G_0^{-1/2}\equiv \mathbbm{1}-\frac{1}{2}\frac{m_Z^2}{\mu^2}\,T+{\cal O}\left(\mu^{-3}\right)
\ee
the kinetic term becomes canonical after a field redefinition $\psi_0\into G_0^{-1/2}\psi_0$, which sends
\be
\wh M_0\into G_0^{-1/2} \wh M_0 G_0^{-1/2}\equiv M_0\,.
\ee
Explicitly,
\be
M_0=\left(\begin{array}{cc} M_1 & \\ & M_2\end{array} \right)-\stb\frac{m_Z^2}{\mu}\;T+\frac{m_Z^2}{\mu^2}\left(\begin{array}{cc}M_1 s^2_w & \frac{M_1+M_2}{2}s_w c_w \\ \frac{M_1+M_2}{2}s_w c_w & M_2 c^2_w\end{array}\right)+{\cal O}(\mu^{-3})\,.
\ee
By diagonalising $M_0$ we recover the well-known approximate masses for wino-like and bino-like neutralinos in the large $\mu$ expansion, e.g.~for $M_2>M_1$,
\be\begin{split}
m_{\chi^0_1}=&\;M_1-\stb s_w^2\frac{m_Z^2}{\mu}-s_w^2\frac{m_Z^2 M_1}{\mu^2}-\frac{s_{2 w}^2\stb^2}{4}\frac{m_Z^4}{\mu^2(M_2-M_1)}+{\cal O}\left(\mu^{-3}\right)\,,\\
m_{\chi^0_2}=&\;M_2-\stb c_w^2\frac{m_Z^2}{\mu}-c_w^2\frac{m_Z^2 M_2}{\mu^2}+\frac{s_{2 w}^2\stb^2}{4}\frac{m_Z^4}{\mu^2(M_2-M_1)}+{\cal O}\left(\mu^{-3}\right)\,.
\end{split}
\ee
The effective Lagrangian for the chargino $\psi_\pm=\frac{1}{\sqrt{2}}\left(\wt W^1\mp i\wt W^2\right)$ is
\be
 {\cal L}_{\rm eff, char}=\psi_+^\dag\,G_\pm\,i\ol\sigma^\mu\partial_\mu\psi_++\psi_-^\dag\,G_\pm\,i\ol\sigma^\mu\partial_\mu\psi_--\psi_+ \wh M_\pm \psi_-\hc
\ee
where
\be
G_\pm=1+c_w^2\frac{m_Z^2}{\mu^2}\,,\qquad \wh M_{\pm}=M_2-c_w^2\stb\frac{m_Z^2}{\mu}
\ee
up to higher-order terms. Canonically normalising the chargino field yields the well-known wino-like chargino mass
\be
m_{\chi^\pm_1}=M_2-\stb c_w^2\frac{m_Z^2}{\mu}-c_w^2\frac{m_Z^2 M_2}{\mu^2}+{\cal O}\left(\mu^{-3}\right)\,.
\ee

A few comments are in order: 
\begin{itemize}
 \item We have integrated out the higgsino fields, which are gauge eigenstates, rather than the corresponding mass eigenstates (higgsino-like neutralinos and charginos). However, the higgsino mixing angles are suppressed by $1/\mu$, so the difference shows up only at higher order.
 \item The expansion breaks down at the degenerate point $M_1=M_2$ because of terms enhanced by $1/(M_2-M_1)$; however, it remains valid unless the difference between the wino and bino mass is far below the electroweak scale. 
 
\end{itemize}
 
Of particular interest for us is the limit where one of the two MSSM Higgs doublets effectively decouples at a scale $\gg m_Z$, while the other is identified with the Standard Model Higgs field $\phi$. It is obtained by replacing $h_u\into s_\beta\phi$ and $h_d\into c_\beta i\sigma^2\phi^*$ in Eq.~\eqref{eq:LeffBW}. The operators affecting the neutralino and chargino masses are then
\be\begin{split}\label{eq:LeffBW2}
{\cal L}_{\rm eff}=\;&\wt W^{a\dag} i\ol\sigma^\mu\partial_\mu\wt W^a+\wt B^\dag i\ol\sigma^\mu\partial_\mu\wt B-\left(\frac{1}{2} M_1\,\wt B\wt B+\frac{1}{2} M_2\,\wt W^a\wt W^a\hc\right)\\
+&\frac{g^2}{2\mu^2}(\phi^\dag\phi)\wt W^{a\dag}i\ol\sigma^\mu\partial_\mu\wt W^a+\frac{{g'}^2}{2\mu^2}(\phi^\dag\phi)\wt B^{\dag}i\ol\sigma^\mu\partial_\mu\wt B+\left(\frac{gg'}{\mu^2}\phi^\dag\tau^a\phi\,\wt W^{a\dag}i\ol\sigma^\mu\partial_\mu\wt B\hc\right)\\
+&\frac{g^2\stb}{2\mu}\phi^\dag\phi\; \wt W^a\wt W^a+\frac{{g'}^2\stb}{2\mu}\phi^\dag\phi\; \wt B\wt B+\frac{gg'\stb}{\mu}\phi^\dag\tau^a\phi\;\wt W^{a}\wt B\hc+\ldots
\end{split}
\ee
The phenomenologically most relevant quantities are the $\chi^0_2-\chi^\pm_1$ mass splitting and the neutralino mixing angle. We observe that, at order $\mu^{-1}$, the masses of both wino-like eigenstates are shifted from $M_2$ by $\stb c_w^2\frac{m_Z^2}{\mu}$ due to the presence of the $\phi^\dag\phi\wt W\wt W$ operator, which after EWSB becomes an $\SU{2}$ invariant wino mass term. There is a similar universal shift at order $\mu^{-2}$ due to the $\SU{2}$ invariant correction to the kinetic term. The mass splitting between  $\chi^0_2$ and $\chi^\pm_1$ is due to wino-bino mass mixing in the neutralino sector, and appears first at ${\cal O}(\mu^{-2})$. In terms of the mixing angle $\theta$, the mass shift is
\be
\Delta_{m_{\chi^\pm_1}-m_{\chi^0_2}}^{\text{mixing}}=(M_2-M_1)\theta^2\,,\qquad \text{where }\theta=\frac{s_{2w}\stb m_Z^2}{2(M_2-M_1)\mu}\,.
\ee
Numerically this is quite small even for only moderately heavy higgsinos and moderate $\tan\beta$. For example, taking $\mu\approx 1$ TeV, $M_2-M_1\approx m_Z$ and $\stb=1$ yields a tree-level mass difference of the order of only 100 MeV, the same order as for the mass splitting induced by electroweak loops (see below).

By replacing the electroweak gauge couplings and factors of $\stb$ in Eq.~\eqref{eq:LeffBW2} by generic Wilson coefficients, and $\mu$ by a generic new physics scale, it is now straightforward to describe a generic non-supersymmetric triplet-singlet system.
The most general triplet-singlet Lagrangian at ${\cal O}(\Lambda^{-1})$ is
\be\begin{split}\label{eq:Leff-nonSUSY}
{\cal L}=\;&i\,\psi^\dag \ol\sigma^\mu D_\mu \psi+ i\,\chi^\dag \ol\sigma^\mu\partial_\mu\chi-\frac{1}{2}\left(M\psi\psi+m\chi\chi\hc\right)\\
&+\left(\frac{1}{2}\frac{\kappa}{\Lambda}\phi^\dag\phi\chi\chi+\frac{1}{2}\frac{\kappa'}{\Lambda}\phi^\dag\phi \psi^a\psi^a+\frac{\lambda}{\Lambda}\phi^\dag\tau^a\phi\;\psi^a\chi\hc\right)\,.
\end{split}
\ee
We assume $M$ and $m$ to be both positive (a relative sign between them would inhibit the ``well-tempering'' mechanism, while general complex phases are beyond the scope of this work). We absorb the $\SU{2}$-invariant corrections to the masses from the $\phi^\dag\phi\chi\chi$ and  $\phi^\dag\phi\psi\psi$ operators into $M$ and $m$. The neutral mass matrix eigenvalues are
\be
m_{\chi^0_1,\,\chi^0_2}=\frac{1}{2}\left(M+m\pm\xi\right)\,\qquad\xi=\sqrt{(M-m)^2+8\frac{\lambda^2 v^4}{\Lambda^2}}
\label{eq:delm-trip}
\ee
with the $\psi^3-X$ mixing angle given by
\be
\sin^2\theta=\frac{1}{2}-\frac{M-m}{2\xi}\,.
\ee
To leading order in the mixing,
\be
\theta=\frac{\sqrt{2}\lambda v^2}{\Lambda(M-m)}\,,
\label{eq:theta-trip}
\ee
with a corresponding mass splitting between the mixed neutral state $m_{\chi^0_2}$ and the pure charged state $m_{\chi^\pm}$
 \be
\Delta_{m_{\chi^\pm}-m_{\chi^0_2}}^{\text{mixing}}=(M-m)\theta^2=\frac{2\,\lambda^2\,v^4}{\Lambda^2(M-m)}\,.
\ee

Besides mixing with the singlet, there are other effects which can potentially split the  masses of the neutral and charged triplet-like mass eigenstates. Note that the naive candidate dimension-5 operator $(\phi^\dag\tau^a\phi)(\psi^b i f^{abc} \psi^c\hc)$ is identically zero by symmetry.
However, at one loop the masses receive corrections from electroweak gauge bosons. The resulting mass splitting in the absence of mixing, i.e.~for $\lambda=0$, is (see e.g.~\cite{Cirelli:2005uq})
\be\label{eq:loopsplitting}
\Delta_{m_{\chi^\pm}-m_{\chi^0_2}}^{\text{one loop}} = \frac{g^2}{16\pi^2}M\left(f\left(\frac{m_W}{M}\right)-c_w^2\,f\left(\frac{m_Z}{M}\right)\right)\,,
\ee
where
\be\label{eq:loopfunction}
f(x)=\frac{x}{2}\left(2\,x^3\log x-2\,x+\sqrt{x^2-4}(x^2+2)\log\frac{x^2-x\sqrt{x^2-4}-2}{2}\right)\,.
\ee
Numerically this varies between  about 150 and 170 MeV for tree-level masses between $10^2$ and $10^4$ GeV.

\section*{Appendix B: The quadruplet-singlet model}
\label{sec:quad}
Even values of $n$ are slightly more complicated since the ${\bf n}$ is a pseudoreal rather than a strictly real representation. We will discuss the example of $n=4$ or the quadruplet model, for which we introduce a Dirac fermion $(\psi,\ol\psi^{\dag})$ transforming in the ${\bf 4}_{\frac{1}{2}}$ as well as a Majorana singlet $\chi$. The Lagrangian is
\be\begin{split}\label{eq:quadrupletL}
{\cal L}=\;&i\,\psi^\dag \ol\sigma^\mu D_\mu \psi+ i\,\ol\psi^{\dag} \ol\sigma^\mu D_\mu \ol\psi+i\,\chi^\dag \ol\sigma^\mu\partial_\mu\chi-\left(M\psi_I\;\ol\psi^I+\frac{1}{2} m\chi\chi\hc\right)\\
&+\left(\frac{1}{2}\frac{\kappa}{\Lambda}\phi^\dag\phi\chi\chi+\frac{\kappa'}{\Lambda}\phi^\dag\phi\psi_I\;\ol\psi^I\hc\right)\\
&+\left(\frac{\zeta_1}{\Lambda}(\phi^\dag\tau^a\phi)(\ol\psi t^a\psi)
-\frac{\zeta_2}{\Lambda}(\phi_i \tau^a{}^i_j\phi^j)(\ol\psi_I t^a{}_J^I\ol\psi^J)
-\frac{\zeta_3}{\Lambda}(\phi^\dag_i\tau^a{}^i_j\phi^{\dag j})(\psi_I t^a{}_J^I\psi^J)\hc\right)\\
&+\left(\frac{\lambda}{\Lambda^2}\;\epsilon_{jl}\;\epsilon_{km}\;\phi^{\dag}_i\phi^l\phi^m\;d_I^{ijk}\chi\ol\psi^I-\frac{\lambda'}{\Lambda^2}\;\epsilon_{kl}\;\phi^{\dag}_i\phi^{\dag}_j\phi^{l}\;d_I^{ijk}\chi\psi^I\hc\right)\,,
\end{split}\ee
with $d_I$ being an orthonormal basis of symmetric $2\times 2\times 2$ tensors containing the Clebsch-Gordan coefficients,
\be\begin{split}
d_1^{ijk}=&\;\delta^i_1\delta^j_1\delta^k_1\,,\quad d_2^{ijk}=\frac{1}{\sqrt{3}}\left(\delta^i_1\delta^j_1\delta^k_2+\delta^i_1\delta^j_2\delta^k_1+\delta^i_2\delta^j_1\delta^k_1\right)\,,\\
\quad d_3^{ijk}=&\;\frac{1}{\sqrt{3}}\left(\delta^i_1\delta^j_2\delta^k_2+\delta^i_2\delta^j_2\delta^k_1+\delta^i_2\delta^j_1\delta^k_2\right)\,,\quad d_4^{ijk}=\delta^i_2\delta^j_2\delta^k_2\,.
\end{split}
\ee
Indices $I,J=1\ldots 4$ are raised and lowered with ${\cal E}=\sigma^1\tensor i\sigma^2$,
\be
({\cal E}_{IJ})=(-{\cal E}^{IJ})=\left(\begin{array}{cccc}&&&1 \\ &&-1& \\ &1&& \\ -1&&&\end{array}\right)\,.
\ee
The quadruplet generators are 
\be\begin{split}
t^1=\left(\begin{array}{cccc}0 & \frac{\sqrt{3}}{2} & 0 & 0 \\ \frac{\sqrt{3}}{2} & 0 & 1 & 0 \\ 0 & 1 & 0 & \frac{\sqrt{3}}{2} \\0 & 0 & \frac{\sqrt{3}}{2} & 0 \\\end{array}\right)\,,&
\quad t^2=\left(\begin{array}{cccc}0 & -\frac{\sqrt{3}}{2}\,i & 0 & 0 \\ \frac{\sqrt{3}}{2}\,i & 0 & -i & 0 \\ 0 & i & 0 & -\frac{\sqrt{3}}{2}i \\0 & 0 & \frac{\sqrt{3}}{2}\,i & 0 \\\end{array}\right)\,,\\
\quad t^3=&\;\left(\begin{array}{cccc}\frac{3}{2} & 0 & 0 & 0 \\ 0 & \frac{1}{2} & 0 &  0 \\ 0 & 0 & -\frac{1}{2} & 0 \\0 & 0 & 0 & -\frac{3}{2} \\\end{array}\right)\,.\end{split}
\ee
Note that the Lagrangian Eq.~\eqref{eq:quadrupletL} contains an additional $\phi^\dag\phi\psi\ol\psi$ operator, leading to $\SU{2}$-breaking mass splittings at the tree level among the quadruplet components after electroweak symmetry breaking. In terms of electric charge eigenstates, $\psi$ and $\ol\psi$ are
\be
(\psi^I)=\left(\begin{array}{l} \psi^{++} \\ \psi^+ \\ \psi^0 \\ \wt \psi^-\end{array}\right)\,,\qquad\qquad (\ol\psi^{J})=\left(\begin{array}{l} \wt\psi^+ \\  \wt\psi^0 \\  \psi^- \\  \psi^{--}\end{array}\right)\,,\qquad\qquad (\ol\psi_J)=\left(\begin{array}{rl}&\!\!\!\!\psi^{--} \\  -&\!\!\!\!\psi^-\\ &\!\!\!\!\wt\psi^0 \\  -&\!\!\!\!\wt \psi^{+}\end{array}\right)\,.
\ee

To find the mass eigenstates we can absorb the $\SU{2}$ invariant contributions to the mass matrix from $\kappa$ and $\kappa'$ into $m$ and $M$. However, the $\zeta_{1,2,3}$ operators, for instance $\ol\psi_I (t^a)^{I}_{J}\;\psi^J$, induce a contribution which is different for doubly charged, singly charged and neutral states. By a further redefinition of $M$ we can assume without loss of generality that this contribution is zero in the neutral sector. The mass of the doubly charged state is then
\be
m_{\chi^{\pm\pm}}=M+\frac{\zeta_1\,v^2}{\Lambda}\,.
\ee
The singly charged mass matrix
\be
{\cal M}_c=\left(\begin{array}{cc} \frac{\sqrt{3}\,\zeta_2\,v^2}{\Lambda} & -M - \frac{\zeta_1\,v^2}{2\,\Lambda} \\ -M + \frac{\zeta_1\,v^2}{2\,\Lambda} & \frac{\sqrt{3}\,\zeta_3\,v^2}{\Lambda} \\\end{array}\right)
\ee
In the limit $\frac{v}{\Lambda}\ll 1$, ${\cal M}_c$ leads to the masses
\be
m_{\chi^\pm_{1,2}} = M \mp 2M\sqrt{\Xi^2 + 3 \Sigma^2} + 6 M \Theta^2 (1-3\Theta^2)\, .
\ee

where
\be
\Xi = \frac{\zeta_1 v^2}{4 M\;\Lambda} \; ,\;\Theta = \frac{(\zeta_3-\zeta_2) v^2}{4 M\;\Lambda} \; ,\; {\rm and} \quad \Sigma = \frac{(\zeta_3+\zeta_2)v^2}{4 M\,\Lambda}\;.
\label{eq:DelSigDef}
\ee
The neutral mass matrix is
\be
{\cal M}_n=\left(\begin{array}{ccc}m & -\frac{1}{\sqrt{3}}\frac{\lambda\,v^3}{\Lambda^2} & -\frac{1}{\sqrt{3}}\frac{\lambda'\,v^3}{\Lambda^2} \\ -\frac{1}{\sqrt{3}}\frac{\lambda\,v^3}{\Lambda^2} & -\dfrac{\zeta_2\,v^2}{\Lambda} & M \\ -\frac{1}{\sqrt{3}}\frac{\lambda'\,v^3}{\Lambda^2} & M & -\dfrac{\zeta_3\,v^2}{\Lambda} \\\end{array}\right)
\ee
Again, for $\frac{v}{\Lambda}\ll 1$, the physical tree-level masses for the neutral states are given, up to possible reordering, by

\be \begin{split}
      m_{\chi^0_1} =&\; m\;(1 + \theta_+^2 + \theta_-^2) -  M\;(\theta_+^2-\theta_-^2)\,,\\ m_{\chi^0_2} =&\; M\,(1+\theta_-^2) + m\,\theta_-^2 + 2M\,(\Sigma + \Theta^2(1-\Theta^2)) \,,\\ m_{\chi^0_3} =&\;  M\,(1+\theta_+^2) - m\,\theta_+^2 - 2M\,(\Sigma - \Theta^2(1-\Theta^2))\,.
 \label{eq:delm-quad}
\end{split}
\ee

with
\be
\theta_{\pm} = \dfrac{(\lambda \pm \lambda')v^3}{\sqrt{6}\;( M \mp m)\Lambda^2} \;.
\label{eq:theta-quad}
\ee

For calculating the precise spectrum we again need to take electroweak corrections into account. One finds
\be
\Delta_{m_{\chi^\pm_{1,2}}-m_{\chi^0_{2,3}}}^{\text{one loop}} = \frac{g^2}{16\pi^2}M\,s_w^2\,f\left(\frac{m_Z}{M}\right)
\ee
and
\be
\Delta_{m_{\chi^{\pm\pm}}-m_{\chi^\pm_{1,2}}}^{\text{one loop}} = \frac{g^2}{16\pi^2}M\,\left((3\,s_w^2-2)\,f\left(\frac{m_Z}{M}\right)+2\,f\left(\frac{m_W}{M}\right)\right)
\ee
with $f(x)$ defined by Eq.~\eqref{eq:loopfunction}.

\section*{Appendix C: The quintuplet-singlet model}
\label{sec:quint}
It is straightforward to generalise the well-tempered triplet-singlet model to any odd $n$. We focus on the simplest example, $n=5$ or the quintuplet-singlet model, whose Lagrangian is
\be
\begin{split}
{\cal L}=\;&i\,\psi^\dag \ol\sigma^\mu D_\mu \psi+ i\,\chi^\dag \ol\sigma^\mu\partial_\mu\chi-\frac{1}{2}\left(M\psi\psi+m\chi\chi\hc\right)\\
&+\left(\frac{1}{2}\frac{\kappa}{\Lambda}\phi^\dag\phi\chi\chi+\frac{1}{2}\frac{\kappa'}{\Lambda}\phi^\dag\phi\psi^A\psi^A+\frac{\lambda}{\Lambda^3}\phi^{\dag i}\phi_j\phi^{\dag k}\phi_\ell\;C_{A\,ik}^{j\ell}  \psi^A\chi\hc\right)
\end{split}
\ee
Here $C_{A\,ik}^{j\ell}$ is the tensor
\be
C_{A\,ik}^{j\ell}=\rho_{A}^{ab}\sigma^{a\,j}_i\sigma^{b\,\ell}_k
\ee
with the $\rho_A^{ab}$ an orthonormal basis of traceless symmetric $3\times 3$ matrices, e.g.
\be\begin{split}
\rho_1=&\;\frac{1}{\sqrt{2}}\left(\begin{array}{ccc}0 & 0 & 0 \\ 0 & 0 & 1 \\ 0 & 1 & 0\end{array}\right)\,,\quad
\rho_2=\frac{1}{\sqrt{2}}\left(\begin{array}{ccc}0 & 0 & 1 \\ 0 & 0 & 0 \\ 1 & 0 & 0\end{array}\right)\,,\quad
\rho_3=\frac{1}{\sqrt{2}}\left(\begin{array}{ccc}0 & 1 & 0 \\ 1 & 0 & 0 \\ 0 & 0 & 0\end{array}\right)\,,\\
\rho_4=&\;\frac{1}{\sqrt{2}}\left(\begin{array}{ccc}1 & 0 & 0 \\ 0 & -1 & 0 \\ 0 & 0 & 0\end{array}\right)\,,\quad
\rho_5=\frac{1}{\sqrt{6}}\left(\begin{array}{ccc}1 & 0 & 0 \\ 0 & 1 & 0 \\ 0 & 0 & -2\end{array}\right)\,.
\end{split}
\ee
In this basis the gauge transformations on the quintuplet are generated by
\be
(t^a)_{AB}=-2i\,\epsilon^{abc}\rho_A^{bd}\rho_B^{cd}\,.
\ee

The mass eigenstates in the quintuplet-singlet system are a doubly charged Dirac fermion $\chi^{++}=\frac{\psi^4-i\psi^3}{\sqrt{2}}$, a singly charged Dirac fermion $\chi^+=\frac{\psi^2-i\psi^1}{\sqrt{2}}$ and two neutral Majorana fermions $\chi^0_{1,2}$, the heavier of which will be $\psi^5$-like if $M>m$. The lighter singlet-like neutral state is our dark matter candidate. Redefining $M\into M-\frac{\kappa' v^2}{\Lambda}$ and  $m\into m-\frac{\kappa v^2}{\Lambda}$, the neutral mass matrix is
\be
M_0=\left(\begin{array}{cc} m & -\sqrt{\frac{2}{3}}\frac{\lambda v^4}{\Lambda^3} \\ -\sqrt{\frac{2}{3}}\frac{\lambda v^4}{\Lambda^3} & M \end{array}\right)\,.
\ee
The mixing angle becomes, to leading order,
\be
\theta=\sqrt{\frac{2}{3}}\frac{\lambda v^4}{\Lambda^3(M-m)}\,,
\label{eq:theta-quint}
\ee
and the neutral-charged mass splitting is once more
\be
\Delta_{m_{\chi^+}-m_{\chi^0_2}}^{\text{mixing}}=\theta^2(M-m)\,.
\label{eq:delm-quint}
\ee

The loop-induced mass splittings between the quintuplet mass eigenstates are given by Eq.~\eqref{eq:loopsplitting} for the difference of the $\chi^+$ and $\chi^0_2$ masses, while
\be
\Delta_{m_{\chi^{++}}-m_{\chi^+}}^{\text{one loop}}=3\;\Delta_{m_{\chi^+}-m_{\chi^0_2}}^{\text{one loop}}\,.
\ee

\newpage
\bibliography{Nuplet}
\bibliographystyle{JHEP-2}

\end{document}